\documentclass[aps,prc,superscriptaddress,showpacs,onecolumn]{revtex4}	
 \usepackage{amsmath}  
 \usepackage{makeidx}
 \usepackage{graphicx}
 \usepackage{amsfonts}
 \usepackage[ansinew]{inputenc}
 \usepackage[usenames,dvipsnames]{pstricks}
 \usepackage{subfigure}
 \usepackage{epsfig}
 \usepackage{pst-grad}
 \usepackage{pst-plot}
 \usepackage{mathrsfs}

              \makeindex
              
\begin{document}
\title{Nuclear matter equation of state including few-nucleon correlations $(A\leq 4)$}
\author{G. R\"{o}pke}
\email{gerd.roepke@uni-rostock.de}
\affiliation {Institut f\"{u}r Physik, Universit\"{a}t Rostock, D-18051 Rostock, Germany}

\date{\today}

\begin{abstract}

Light clusters (mass number $A \leq 4$) in nuclear matter at subsaturation densities are described using a quantum statistical 
approach. In addition to self-energy and Pauli-blocking, 
effects of continuum correlations are taken into account to 
calculate the quasiparticle properties and abundances of light elements. 
Medium-modified quasiparticle properties are important ingredients to derive a nuclear matter equation of state 
applicable in the entire region of warm dense matter below saturation density.
Moreover, the contribution of continuum states to the equation of state is considered.
The effect of correlations within the nuclear medium on the quasiparticle energies is estimated.
The properties of light 
clusters and continuum correlations in dense matter are of interest for nuclear structure calculations, heavy ion collisions, 
and for astrophysical applications such as the formation of neutron stars in core-collapse supernovae. 
 
\end{abstract}


\maketitle

 \section{Introduction}
\label{Sec:Introduction}
 
We investigate nuclear matter in thermodynamic equilibrium, confined in the volume $\Omega$ at temperature $T$. 
It consists of $N_n$ neutrons
(total neutron density $n^{\rm tot}_n=N_n/\Omega$)
and $N_p$ protons (total proton density $n^{\rm tot}_p=N_n/\Omega$).
We are interested in the subsaturation region where the baryon density 
$n_B= n^{\rm tot}_n+n^{\rm tot}_p \leq n_{\rm sat}$
with the saturation density $n_{\rm sat} \approx 0.16$ fm$^{-3}$, the temperature $T \leq 20$ MeV, 
and proton fraction $Y_p=n^{\rm tot}_p/n_B$ between 0
and 1. This region of warm dense matter is of interest not only for nuclear structure calculations 
and heavy ion collisions 
explored in laboratory experiments \cite{Natowitz}, but also in astrophysical applications. 
For instance,  core-collapse supernovae at post-bounce stage are evolving within this region of the phase space \cite{Tobias}, 
and different processes such as  neutrino emission and absorption, which strongly depend on the composition of warm dense matter,
influence the mechanism of core-collapse supernovae. 
 
In particular, the standard versions \cite{LS,Shen} of the nuclear matter equation of state (EOS) 
for astrophysical simulations have been improved recently, see
\cite{SR,Arcones2008,Armen2009,Typel,Gulminelli2010,NSE,Hempel,Furusawa,shenTeige,Providencia,Avancini,Jaqaman,Vergleich,Hempel2013,Gulminelli2013,Gulminelli2014,NSETabellen,Hempel2014}. 
Here, we will not discuss different approaches but rather contribute to a special question, 
the treatment of light clusters which is a long-standing problem \cite{RMS}. 
A simple chemical equilibrium of free nuclei is not applicable up to saturation density 
because medium modifications by self-energy shifts and Pauli blocking become relevant.
Concepts such as the heuristic excluded volume approach or in-medium nuclear cluster energies 
within the extended Thomas-Fermi approach may be applied 
to heavier clusters but are not satisfactory to describe light clusters that require a more fundamental quantum statistical (QS) approach.
 
 We consider the total proton number density
$n^{\rm tot}_p$, the total neutron number density $n^{\rm tot}_n$, and the temperature $T$ as independent thermodynamic variables. 
Weak interaction processes leading to $\beta$ equilibrium are not considered. The chemical potentials $ \mu_n,\mu_p $ are an alternative 
to $n^{\rm tot}_p$ and $n^{\rm tot}_n$ in characterizing thermodynamic equilibrium of warm dense matter.
The relations
\begin{equation}
\label{eos0}
\frac{1}{\Omega}N_n=n^{\rm tot}_n(T,\mu_n,\mu_p),\qquad \frac{1}{\Omega}N_p=n^{\rm tot}_p(T,\mu_n,\mu_p)
\end{equation}
are equations of state that relate the set of thermodynamic quantities $\{T,\mu_n,\mu_p\}$ to $\{T,n^{\rm tot}_n,n^{\rm tot}_p\}$.
We give solutions for these EOS for warm dense matter. Further thermodynamic variables are consistently derived after a thermodynamic potential
is found by integration, see App. \ref{App:TD}.

To treat the many-nucleon system (nuclei and nuclear matter) at densities up to saturation,
semi-empirical mean-field approaches have been worked out. Based on 
the Hartree-Fock-Bogoliubov approximation and related quasiparticle concepts 
such as the Dirac-Brueckner-Hartree-Fock (DBHF) approach for the nuclear matter EOS, see  \cite{Klahn:2006ir},
semi-empirical approaches such as the Skyrme parametrization \cite{Skyrme} or relativistic mean-field (RMF) approaches give an adequate description 
of the properties of nuclear matter near the saturation density. For a discussion of different versions of these models see, 
for instance, Ref.~ \cite{Providencia,Avancini,Hempel2014}.
The mean-field potentials may be considered as density functionals that include various correlations, 
beyond a microscopic Hartree-Fock-Bogoliubov approximation. 
In this work we use the DD-RMF parametrization according to Typel, Refs. \cite{Typel2005,Typel1999}. 
Other parametrizations of the nucleon quasiparticle energies can alternatively be used  
to optimize the description of nuclear matter near saturation density.

For strongly interacting nuclear matter considered here, in particular warm dense matter in the low-density region, 
correlations are important so that a simple mean-field description is not satisfactory. 
A quantum statistical (QS) approach \cite{FW,AGD}, see Sec. \ref{sec:Green}, can treat the many-particle aspects in a systematic way, 
using the methods of thermodynamic Green functions, diagram techniques, or path integral methods. 
A signature of strong correlations is the formation of bound states.
In the low-density limit, we can consider the many-nucleon system as an ideal mixture of clusters (nuclei) 
where the interaction is reduced to accidental 
(reacting) collisions, leading to chemical equilibrium as given by the mass-action law. 
This so-called nuclear statistical equilibrium (NSE), see \cite{NSE}, has several shortcomings, such as the exclusion of excited states, 
in particular continuum correlations,
and the failure to account for the interaction between the different components (single nucleons as well as nuclei) 
that is indispensable when  approaching saturation density. 
Both problems are discussed in the present work. 

To describe correlations in warm dense matter we restrict our treatment to light elements: deuteron $d$ ($^2$H), triton $t$ ($^3$H), 
helion $h$ ($^3$He), and  $\alpha$ ($^4$He), in addition to free neutrons ($n$) and protons ($p$).
The QS approach can be extended to describe further clusters with $A > 4$, see \cite{Debrecen}, 
but is not well worked out for this regime until now.
An alternative approach to include heavy nuclei is the concept of the excluded volume (EV), see \cite{Hempel}. In a simple semi-empirical
approach, the effect of Pauli blocking is replaced by the strong repulsion determined by the excluded volume. 
The comparison between the EV model and the QS approach \cite{HempelRoepke} gives qualitatively similar results, 
although in the EV model the center of mass motion of clusters is not systematically treated (for instance, effective mass and 
quantum condensation effects), the light clusters such as the deuteron are not well described by a hard core potential, 
not depending on the energy, and correlations in the continuum are not considered. 
Here, we present a theory for nuclear systems under conditions where clusters with $A > 4$ are irrelevant, 
(see Refs.  \cite{Shen,Furusawa,shenTeige,Vergleich} for an illustration of the parameter space $\{T, n_B, Y_p\}$ where such regions are shown).
Heavier clusters can be included as e.g. done in Ref. \cite{SR}, and the combination with the EV model to treat heavier nuclei will be
discussed in future work.

Our aim is to describe nuclear matter in the entire region of subsaturation densities, 
connecting the single-nucleon quasiparticle 
approach that reproduces the properties near $n_{\rm sat}$ to the low-density limit where a cluster-virial expansion
\cite{clustervirial} is possible. 
This is achieved by considering the constituents $c=\{d,t,h,\alpha\}$ as quasiparticles
with energy $E_c(P; T, n^{\rm tot}_n,n^{\rm tot}_p)$ which depends not only of  the center of mass momentum $P$, 
but also the set of thermodynamic parameters $\{T, n^{\rm tot}_n,n^{\rm tot}_p\}$ or, according to the EOS (\ref{eos}), $\{T, \mu_n,\mu_p\}$. 
These medium-dependent quasiparticle cluster energies are obtained 
from an in-medium few-body wave equation derived within a Green-function approach, see Sec. \ref{Sec:Few-body}. 
More details including fit formulae for the cluster quasiparticle energies are given in \cite{R,R2011}.
We account for the contributions of self-energy and Pauli blocking
to the quasi-particle energies that describe the light elements moving in warm dense matter.
The Coulomb energy that is screened in dense matter can be omitted for $Z \leq 2$.

The treatment of excited states and continuum correlations leads to a virial expansion \cite{Huang} that describes rigorously the low-density limit.
The generalized Beth-Uhlenbeck approach \cite{SRS} allows to implement both limits, the low-density NSE
 with the virial expansion as benchmark \cite{HS}, and the behavior near saturation density,  where the 
 quasi-particle concept for the nucleons is applicable. The low density region, where the contribution of the continuum correlations to the 
 virial expansion has to be taken into account, was investigated recently in context with the generalized RMF approach \cite{VT}. 
 The bound states (light elements) gradually disappear due to Pauli blocking before saturation density is reached.
 Introducing the quasiparticle concept, we have to be careful to avoid double counting because part of the continuum correlations 
 is already implemented in the quasiparticle energy shift, see \cite{clustervirial}. 
 We give expressions for the remaining residual continuum correlations that are extrapolated to higher densities in Sec. \ref{sec:Virial}.
 
 Another problem is the treatment of correlations in the medium. Although the formalism has been worked out \cite{RMS2,cmf,schuckduk}, 
 the resulting cluster mean-field 
 equations have not yet been solved in a self-consistent way. Comparing with the ideal Fermi distribution of free nucleons, 
 the occupation of the phase space is changed if correlations in the medium are taken into account. 
 We propose a simple parametrization to improve on the Fermi distribution of the ideal nucleon gas. 
 Results are given in Sec. \ref{Sec:Results}, and some general issues that need to be resolved to devise an improved
 EOS are discussed in the final Sec. \ref{Sec:Discussion}.

\section{Green functions approach and quasiparticle concept}
\label{sec:Green}

\subsection{Cluster decomposition of the equation of state}

The nuclear matter EOS ($ \Omega$ is the system volume, $\tau = \{n,p\}$)
\begin{equation}
\label{EOS}
  n^{\rm tot}_\tau(T,\mu_n,\mu_p)=\frac{1}{\Omega}\sum_{p_1,\sigma_1} \int \frac{d \omega}{2 \pi} \frac{1}{e^{(\omega-\mu_\tau)/T}+1}
  S_\tau(1,\omega)
\end{equation}
is obtained from the spectral function $S_\tau(1,\omega;T,\mu_n,\mu_p)$
which is related to the self-energy, see \cite{FW,AGD}:
\begin{equation}
\label{spectral}
 S_\tau(1,\omega) = \frac{2 {\rm Im}\Sigma(1,\omega-i0)}{ 
(\omega - E(1)- {\rm Re} \Sigma(1,\omega))^2 + 
({\rm Im}\Sigma(1,\omega-i0))^2 }\,.
\end{equation}
The single-nucleon quantum state $|1\rangle$
can be chosen as $1 = \{{\bf p}_1, \sigma_1,\tau_1\}$ which denotes wave number, spin, and isospin, respectively.
The EOS (\ref{EOS}) relates the total nucleon numbers $N^{\rm tot}_\tau$ or the particle densities $n^{\rm tot}_\tau$ to the 
chemical potentials $\mu_\tau$ of neutrons or protons so that we can switch from the densities to the chemical potentials. 
On the other hand, if this EOS is known in some approximation, all other thermodynamic quantities are obtained consistently
after changing over to a thermodynamic potential, see App. \ref{App:TD}.

The spectral function $S_\tau(1,\omega;T,\mu_n,\mu_p)$ and the corresponding correlation functions are quantities, well-defined in the grand canonical
ensemble characterized by $\{T,\mu_n,\mu_p\}$. 
The self-energy $\Sigma(1,z;T,\mu_n,\mu_p)$ depends, besides the single-nucleon quantum state $|1\rangle$, on the complex frequency $z$
and is calculated at the Matsubara frequencies.
Within a perturbative approach it can be represented by Feynman diagrams. A cluster decomposition 
with respect to different few-body channels ($c$) is possible, characterized, for instance, by the nucleon number $A$, 
as well as spin and isospin variables.

Using the cluster decomposition of the self-energy which takes into account, in particular, cluster formation,
we obtain
\begin{eqnarray}
\label{eos}
&&  n^{\rm tot}_n(T,\mu_n,\mu_p)= \frac{1 }{ \Omega} \sum_{1}\int \frac{d \omega}{2 \pi}
f_{1,0}(\omega) S_n(1,\omega)= \frac{1 }{ \Omega} \sum_{A,\nu,P}N 
f_{A,Z}[E_{A,\nu}(P;T,\mu_n,\mu_p)] , \nonumber\\ 
&&  n^{\rm tot}_p(T,\mu_n,\mu_p)= \frac{1 }{ \Omega} \sum_{1}\int \frac{d \omega}{2 \pi}
f_{1,1}(\omega) S_p(1,\omega)= \frac{1 }{ \Omega} \sum_{A,\nu,P}Z 
f_{A,Z}[E_{A,\nu}(P;T,\mu_n,\mu_p)] \, ,
\label{quasigas}
\end{eqnarray}
where $\bf P$ denotes the center of mass (c.o.m.) momentum of the cluster (or, for $A=1$, the momentum of the nucleon). 
The internal quantum state $\nu$ contains the proton number $Z$ and neutron number $N=A-Z$ of the cluster, 
\begin{equation}
f_{A,Z}(\omega;T,\mu_n,\mu_p)=\frac{1}{ \exp [(\omega - N \mu_n - Z \mu_p)/T]- (-1)^A}
\label{vert}
\end{equation}
is the Bose or Fermi distribution function for even or odd $A$,
respectively, that is depending on $\{T,\mu_n,\mu_p\}$. 
The integral over $\omega$ is performed within the quasiparticle approach, the quasiparticle energies $E_{A,\nu}(P;T,\mu_n,\mu_p)$
are depending on the medium characterized by $\{T,\mu_n,\mu_p\}$. These in-medium modifications will be detailed in the following sections 
\ref{sec:cmf}, \ref{mediummodi}, and \ref{Sec:Few-body}.

We analyze the contributions of the clusters ($A\geq 2$), suppressing the thermodynamic variables $\{T,\mu_n,\mu_p\}$.
We have to perform the integral over the c.o.m. momentum $\bf P$ what, in general, must be done numerically since the dependence of the 
in-medium quasiparticle energies $E_{A,\nu}(P;T,\mu_n,\mu_p)$ on  $\bf P$ is complex.
The summation over $\nu$ concerns the bound states as far as they exist, as well as the continuum of scattering states. 
Solving the few-body problem what is behind the $A$-nucleon T matrices in the Green function approach, 
we can introduce different channels ($c$) characterized, e.g., by spin and isospin quantum numbers. 
This intrinsic quantum number will be denoted by $\nu_c$, and we have
 in the non-degenerate case $\left[ \,\sum_P \to \Omega/(2 \pi)^3 \int d^3P \,\right]$
\begin{eqnarray}
\label{components}
&&\frac{1 }{ \Omega} \sum_{\nu,P}f_{A,Z}[E_{A,\nu}(P)]=\sum_{c} e^{\left(N\mu_n+Z\mu_p\right)/T} 
 \int \frac{d^3 P}{(2 \pi)^3}\sum_{\nu_c}g_{A,\nu_c} e^{-E_{A,\nu_c}(P)/T} =\sum_{c}
 \int \frac{d^3 P}{(2 \pi)^3} z^{\rm part.}_{A,c}(P)
\end{eqnarray}
with  $g_{A,c} =2 s_{A,c} +1$  the degeneration factor in the channel $c$.
The partial density of the channel $c$ at $\bf P$
\begin{equation}
\label{zpart}
 z^{\rm part.}_{A,c}(P;T,\mu_n,\mu_p)=e^{\left(N\mu_n+Z\mu_p\right)/T}\left\{\sum_{\nu_c}^{\rm bound}g_{A,\nu_c} e^{-E_{A,\nu_c}(P)/T} 
 \Theta\left[-E_{A,\nu_c}(P)+E_{A,c}^{\rm cont}(P)\right]+z^{\rm cont}_{A,c}(P)\right\}\,
\end{equation}
contains the intrinsic partition function. It can be decomposed in the bound state contribution and the contribution of scattering states $z^{\rm cont}_{A,c}(P;T,\mu_n,\mu_p) $.

The summations of (\ref{zpart}) over $A, c$ and $\bf P$ remain to be done for the EOS (\ref{eos}), and $Z$ may be included in $c$.
The region in the parameter space, in particular $\bf P$, where bound states exist, may be restricted what is expressed by 
the step function $\Theta(x)=1, x \geq 0;\,\,=0$ else. The continuum edge of scattering states is denoted by 
$E_{A,c}^{\rm cont}(P;T,\mu_n,\mu_p)$, see Eq. (\ref{Econt}) below.

For instance,
in the case $A=2$ the deuteron is found in the spin-triplet, isospin-singlet channel ($A,c\to d$) as bound state. In addition,
in the same channel we have also contributions from scattering states, i.e. continuum contributions, characterized by the 
relative momentum as internal quantum number. 
Thus, in the isospin-singlet (spin-triplet) channel of the two-particle case ($A=2$) we have in the zero density limit 
the deuteron as bound state [$g_d=3$, $E_{2,T_I=0}(P)=E_d(P)=E_d^0 + \hbar^2 P^2/4m$, $E_d^0=-B_d=-2.225$ MeV], 
and according to the Beth-Uhlenbeck formula \cite{Huang}, see \cite{RMS},
\begin{equation}
\label{B-U}
\int \frac{d^3 P}{(2 \pi)^3} z^{\rm part.}_{d}(P)=
\int \frac{d^3 P}{(2 \pi)^3}\left[\sum_{\nu_c}^{\rm bound}g_{A,\nu_c} e^{-E_{A,\nu_c}(P)/T}+z^{\rm cont}_{A,c}(P)
 \right]=\frac{2^{3/2}}{\Lambda^3} \left[g_d e^{-E_d^0/T}+\int_0^\infty \frac{dE}{\pi} e^{-E/T}\frac{d}{d E} \delta_{2,T_I=0}^{\rm tot}(E) \right]
\end{equation}
with
$\Lambda=(2 \pi \hbar^2/m T)^{1/2}$ being the baryon thermal wavelength (the neutron and proton mass are approximated by 
$m_\tau \approx m=939.17$ MeV$/c^2$), and $  \delta_{2,T_I=0}^{\rm tot}(E)= \sum_{S,L,J}(2J+1) \delta_{{^{2S+1}}L_J}(E) $ 
the isospin-singlet ($T_I=0$) scattering phase shifts with angular momentum $L$ 
as function of the energy $E$ of relative motion. A similar expression can also be derived for the isospin-triplet channel (e.g. two neutrons) 
where, however, no bound state occurs, see also \cite{HS} where detailed numbers are given.

At this point we mention that the NSE is recovered if the summation over $\nu_c$ is restricted to only the bound states 
(nuclei), neglecting the contribution of correlations in the continuum. Furthermore, for $E_{A,\nu_c}(P;T,\mu_n,\mu_p)$
the bound state energies of the isolated nuclei are taken, neglecting the effects of the medium such as mean-field 
terms or contributions due to correlations in the medium. Both aspects will be investigated in the present work. 
In contrast to the semi-empirical treatment of the medium contribution to the EOS within the excluded volume concept, see \cite{Hempel}, 
we use the Pauli blocking terms within a systematic quantum statistical approach.

Note that the subdivision (\ref{zpart}), (\ref{B-U}) into a bound state contribution and a contribution 
$z^{\rm cont}_{A,c}(P;T,\mu_n,\mu_p)$ of 
continuum states is not unique. We will use another decomposition which follows after performing an integration by parts, 
see Eq.~(\ref{intrd}) below.

\subsection{Cluster mean-field approximation}
\label{sec:cmf}

To go to finite densities, the main problem is the medium modification of few-body properties which defines also 
$ E_{A,\nu_c}(P;T,\mu_n,\mu_p) $ in the contribution (\ref{components}) of the different components.
Quasiparticles are introduced considering the propagation of few nucleon cluster (including $A=1$) in warm dense matter.
The Green function approach describes the propagation of a single nucleon by a Dyson equation governed by the self-energy,
and the few-particle states are obtained from a Bethe-Salpeter equation containing the effective interaction kernel.
Both quantities, the effective interaction kernel and the single-particle self-energy, should be approximated consistently.
Approximations which take cluster formation into account have been worked out \cite{RMS2,cmf,clustervirial}, the cluster 
mean-field approximation is outlined in App. \ref{App:CMF}.

For the $A$-nucleon cluster, the  in-medium Schr\"odinger equation 
\begin{eqnarray}
&&[E_{\tau_1}(p_1;T,\mu_n,\mu_p)+\dots + E_{\tau_A}(p_A;T,\mu_n,\mu_p) - E_{A \nu}(P;T,\mu_n,\mu_p)]\psi_{A \nu P}(1\dots A)
\nonumber \\ &&
+\sum_{1'\dots A'}\sum_{i<j}[1-n(i;T,\mu_n,\mu_p)- n(j;T,\mu_n,\mu_p)]V(ij,i'j')\prod_{k \neq 
  i,j} \delta_{kk'}\psi_{A \nu P}(1'\dots A')=0\,
\label{waveA}
\end{eqnarray}
is derived from the Green function approach.
This equation contains the effects of the medium in the single-nucleon quasiparticle shift 
\begin{equation}
\label{selfe}
\Delta E_{\tau}^{\rm SE}(p;T,\mu_n,\mu_p) = E_\tau(p;T,\mu_n,\mu_p)-\sqrt{ m^2 c^4+\hbar^2 c^2 p^2} 
+ m c^2 \approx  E_\tau(p;T,\mu_n,\mu_p)-\frac{\hbar^2p^2}{2 m} 
\end{equation}
(nonrelativistic case), as well as in the Pauli blocking terms given by the occupation numbers $n(1;T,\mu_n,\mu_p)$
in the phase space of single-nucleon states $|1 \rangle \equiv |{\bf p}_1,\sigma_1,\tau_1 \rangle$.  
Thus, two effects have to be considered, the quasiparticle
energy shift and the Pauli blocking. 

In the lowest order of perturbation theory with respect to the nucleon-nucleon interaction $V(12,1'2')$, 
the influence of the medium on the few-particle states ($A=1 \dots 4$) 
is given by the Hartree-Fock shift
\begin{equation}
\label{mHF}
\Delta E^{\rm HF}_{\tau_1}(p_1)=\sum_2 V(12,12)_{\rm ex}f_{1,\tau_2}(2)
\end{equation}
 and, consistently, 
the Pauli blocking terms 
\begin{equation}
\label{mPb}
\Delta V^{\rm Pauli}_{12}(12,1'2')=-\frac{1}{2} \left[f_{1,\tau_1}(1)+f_{1,\tau_{1'}}(1')\right]V(12,1'2') 
\end{equation}
 for $A=2 \dots 4$, see (\ref{cSE}),  (\ref{cPb}) neglecting the contributions with $B >1$.

Both terms have a similar structure, besides the nucleon-nucleon interaction $V$ the single-nucleon Fermi distribution 
$f_{1,\tau_1}(1)=f_{1,n/p}[E(p_1);T,\mu_n,\mu_p]$ occurs. In this simplest approximation, only the free nucleons contribute to 
the self-energy shift and the Pauli blocking. The distribution function is the Fermi distribution with the parameter set $\{T, \mu_n, \mu_p\}$.

It is obvious that also the nucleons found in clusters contribute to the mean field leading to the self-energy, 
but occupy also phase space and contribute to the Pauli blocking. 
The cluster mean-field approximation which considers also the few-body T matrices in the self-energy 
and in the kernel of the Bethe-Salpeter equation leads to similar expressions (\ref{mHF}), (\ref{mPb}) 
but the free-nucleon Fermi distribution
$f_{1,\tau_1}(1)$ replaced by the effective occupation number (\ref{effocc})
\begin{equation}
\label{nnorm}
n(1) = f_{1,\tau_1}(1) + \sum^\infty_{B=2} \sum_{\bar \nu, \bar P} \sum_{2 \dots B}
B f_B\left[E_{B , \bar \nu}(  \bar P;T,\mu_n,\mu_p)\right] |\psi_{B \bar \nu \bar P}(1 \dots B)|^2\,,
\end{equation}
with contains also the distribution function $f_B[E_{B, \bar \nu}(  \bar P)]$ for the abundance of the 
different cluster states 
and the respective wave functions $\psi_{B \bar \nu \bar P}(1 \dots B)$. (The variable $Z$ has not been given explicitly.)
For the quantum statistical derivation see the references given in App. \ref{App:CMF}.

Because the self-consistent determination of $n(1;T,\mu_n,\mu_p)$ for given $\{T, \mu_n, \mu_p\}$ is very cumbersome,
we consider appropriate approximations. In particular, we use the Fermi distribution with new
parameters $\{T_{\rm eff},\mu_n^{\rm eff}, \mu_p^{\rm eff}\}$ (effective temperature and chemical potentials),
\begin{equation}
\label{effparameter}
n(1;T,\mu_n,\mu_p)\approx f_{1,\tau_1}(1; T_{\rm eff},\mu_n^{\rm eff}, \mu_p^{\rm eff})\,.
\end{equation}
These effective parameters allow to reproduce some moments of the occupation number distribution. 
For instance, besides the normalization (total neutron/proton number)
\begin{equation}
\label{moment1}
 \sum_{p_1} n(1;T,\mu_n,\mu_p)=N^{\rm tot}_{\sigma_1,\tau_1}=\sum_{p_1} f_{1,\tau_1}(1;T_{\rm eff},\mu_n^{\rm eff}, \mu_p^{\rm eff})
\end{equation}
also
\begin{equation}
\label{moment2}
 \sum_{1} p_1^2 n(1;T,\mu_n,\mu_p)=\sum_{1} p_1^2  f_{1,\tau}(1;T_{\rm eff},\mu_n^{\rm eff}, \mu_p^{\rm eff})\,
\end{equation}
can be used to fix the values of  $T_{\rm eff},\mu_n^{\rm eff}, \mu_p^{\rm eff}$ as functions of $\{T,\mu_n,\mu_p\}$.

This ansatz contains the special case where the medium is described by  non-interacting single-nucleons states.
Then, the effective parameter values coincide with $T,\mu_n,\mu_p$. We come back to the parametrization, Eq. (\ref{effparameter}),
  where $T_{\rm eff},\mu_n^{\rm eff}, \mu_p^{\rm eff} $ 
are functions of $\{T,\mu_n,\mu_p\}$ according to (\ref{moment1}) and (\ref{moment2}),  below in Sec. \ref{sec:effT}.

\subsection{Medium modification of few-body properties in warm dense matter }
\label{mediummodi}

For the $A$-nucleon cluster, the  in-medium Schr\"odinger equation (\ref{waveA}) is derived, depending on the occupation numbers 
$n(i;T,\mu_n,\mu_p)$ of the single-nucleon states $|i\rangle$. As a consequence, the solutions 
(the energy eigenvalues $E_{A \nu}(P;T,\mu_n,\mu_p)$ and the wave functions) 
will also depend on the parameters which characterize the occupation numbers. 

We change the parameter introducing new variables $\{T_{\rm eff},\mu_n^{\rm eff}, \mu_p^{\rm eff}\}$ according (\ref{effparameter}) to characterize the 
occupation number distribution. We calculate the eigenvalues as function of these new variables that are related to the original variables 
$\{T,\mu_n,\mu_p\}$. We go a step further and switch from the chemical potentials to densities so that we use $\{T_{\rm eff}, n_B, Y_p\}$ as
variables to characterize the occupation number distribution,
\begin{equation}
\label{effparameter1}
n(1;T,\mu_n,\mu_p)\approx \tilde f_{1,\tau_1}(1; T_{\rm eff},n_B,Y_p)\,.
\end{equation}
The tilde $\tilde f_{1,\tau}(1; T_{\rm eff},n_B,Y_p)$ denotes a Fermi distribution as a function of densities instead the chemical potentials.
Implicitly, $\tilde f_{1,\tau}(1)$ depends on the effective parameter values $\{T_{\rm eff},\mu_n^{\rm eff}, \mu_p^{\rm eff}\}$ so that the 
normalization holds, i.e. $\mu_\tau^{\rm eff}(T_{\rm eff},n_B,Y_p)$ are the solutions of the normalization conditions (\ref{moment1}).

The use of $n_B=n_n^{\rm tot}+n_p^{\rm tot}$ and $Y_p= n_p^{\rm tot}/n_B$ realizes that all nucleon participate in the phase space occupation 
[see (\ref{nnorm})], and $T_{\rm eff}$ is 
a further parameter the takes into account the formation of correlations in the medium, see Eq. (\ref{Teff}) given below. 
As a consequence,   the solutions 
(the energy eigenvalues $E_{A \nu}(P;T, n_B, Y_p,T_{\rm eff})$ and the wave functions) 
will also depend on the parameters  $\{T_{\rm eff}, n_B, Y_p\}$ which now characterize the occupation numbers, but are, in principle, 
functions of $\{T,\mu_n,\mu_p\}$. 

To evaluate the dependence of the cluster energy eigenvalues $E_{A \nu}(P;T, n_B, Y_p,T_{\rm eff})$ on $\{T, n_B, Y_p,T_{\rm eff}\}$
we  solve the in-medium Schr\"odinger equation
\begin{eqnarray}
&&\left[E_{\tau_1}(p_1;T,n_B,Y_p)+\dots + E_{\tau_A}(p_A;T,n_B,Y_p) - E_{A \nu}(P;T,n_B,Y_p,T_{\rm eff})\right]\psi_{A \nu P}(1\dots A)
\nonumber \\ &&
+\sum_{1'\dots A'}\sum_{i<j}\left[1- \tilde f_{1,{\tau_i}}(i;T_{\rm eff},n_B,Y_p)- \tilde f_{1,{\tau_j}}(j;T_{\rm eff},n_B,Y_p)\right]
V(ij,i'j')\prod_{k \neq 
  i,j} \delta_{kk'}\psi_{A \nu P}(1'\dots A')=0\,
\label{waveAfree}
\end{eqnarray}
obtained from  (\ref{waveA}) replacing the occupation numbers $n(i;T,\mu_n,\mu_p)$ by a Fermi distribution 
$ \tilde f_{1,{\tau_i}}(i;T_{\rm eff},n_B,Y_p)$.
This equation contains the effects of the medium in the single-nucleon quasiparticle shift 
$\Delta E_{\tau_1}^{\rm SE}(P;T,n_B,Y_p) $, Eq. (\ref{selfe}),
as well as in the Pauli blocking terms given by the occupation numbers $ \tilde f_{1,{\tau_i}}(i;T_{\rm eff},n_B,Y_p)$
in the phase space of single-nucleon states $|i \rangle$. 

Obviously the bound state wave functions and energy eigenvalues $E_{A \nu}(P;T,n_B,Y_p,T_{\rm eff})$ as
well as the scattering phase shifts become dependent on the effective temperature $T_{\rm eff}$ and the
densities $n_\tau^{\rm tot}$. In particular, we obtain the cluster quasiparticle shifts 
\begin{equation}
\label{qushift}
E_{A,\nu}(P)-E^0_{A,\nu}(P)
=\Delta E_{A,\nu}^{\rm SE}(P)+\Delta E_{A,\nu}^{\rm Pauli}(P) 
+\Delta E_{A,\nu}^{\rm Coulomb}(P) 
\end{equation}
with the free contribution $ E^0_{A,\nu}(P)=E^0_{A,\nu}+\hbar^2 P^2/(2Am) $.
Expressions for the in-medium self-energy shift $\Delta E_{A,\nu}^{\rm SE}(P;T,n_B,Y_p) $ and Pauli blocking 
$ \Delta E_{A,\nu}^{\rm Pauli}(P;T_{\rm eff},n_B,Y_p) $ 
are given in Sec.~\ref{sec:shiftsoflight} and App. \ref{App:qushift} below. 
We added the Coulomb shift due to screening effects which can be approximated by the Wigner-Seitz expression. 
For the light elements with $Z \leq 2$ considered here, the Coulomb corrections are small compared with the 
other contributions and are omitted. 

Of special interest are the binding energies 
\begin{equation}
\label{Ebind}
B^{\rm bind}_{A,\nu}(P;T,n_B,Y_p,T_{\rm eff})
=-[ E_{A,\nu}(P;T,n_B,Y_p,T_{\rm eff})-E^{\rm cont}_{A,\nu}(P;T,n_B,Y_p)]
 \end{equation}
with
\begin{eqnarray}
\label{Econt}
E^{\rm cont}_{A,\nu}(P;T,n_B,Y_p)
&=&N E_n(P/A;T,n_B,Y_p)+Z E_p(P/A;T,n_B,Y_p),
\end{eqnarray}
that indicate the energy difference between the bound state and the continuum of free (scattering) states at the same total momentum $P$. 
This binding energy determines the yield of the different nuclei 
according to Eq.~(\ref{quasigas}), where 
the summation over $P$ is restricted to that region where bound states exist, i.e. $B^{\rm bind}_{A,\nu}(P;T,n_B,Y_p,T_{\rm eff}) \ge 0$. 

In addition,  the continuum states solving Eq. (\ref{waveAfree}) are also influenced by the medium effects, but the results are less obvious.
We give some estimations in Sec. \ref{sec:Virial}.

\section{Quasiparticle contributions to the EOS}
\label{Sec:Few-body}

We analyze the contributions of different mass numbers $A$ to the EOS (\ref{eos}). The single-nucleon contribution $A=1$ is extensively
discussed, the quasiparticle picture is well elaborated and broadly applied. An exhaustive discussion of the two-nucleon contribution ($A=2$) 
has been given in Ref. \cite{SRS}. Besides the bound state 
part, also the scattering states have been treated. A generalized Beth-Uhlenbeck equation has been considered 
where not only the low-density limit (second virial coefficient) is correctly reproduced, but also the mean-field terms are consistently
included avoiding double counting. The results can be applied to finite densities of warm dense matter up to saturation density. 
Correlations in the medium have been neglected so that the effective occupation numbers (\ref{effparameter}) 
are approximated by $\tilde f_{1,\tau}(1; T,n_B,Y_p)$.

We are interested in including all light clusters ($A \leq 4$). Besides the cluster-virial expansion in the low-density limit, the medium 
modifications raising up with increasing density are of interest, and the effect of correlations in the medium is discussed. 
The behavior of bound states has been investigated in previous work \cite{R,R2011}, and some usable results are available, 
but the continuum contributions remain until now very difficult to treat.
We give some estimations and simple interpolation formulae.

In the present work, the sum over $A$ will be restricted to $1\leq A \leq 4$ (light elements).
The contribution of heavier clusters $A > 4$ to the EOS is not considered in calculating the EOS. This confines the 
region of thermodynamic parameters to that region in the phase space where heavier clusters are not of relevance, see, e.g., 
\cite{Shen,Furusawa,shenTeige,Vergleich}.
Heavier clusters may be included, however, different considerations (such as the excluded volume concept) in addition 
to the NSE have to be performed to treat higher densities. Besides some QS calculations \cite{Debrecen}, mostly 
the excluded volume concept, see \cite{LS,Shen,Hempel}, 
is used to give a semi-empirical treatment of the medium-modified contribution of heavier clusters to the EOS.

\subsection{Single-nucleon quasiparticle approximation}

Before improving the low-density limit of the EOS considering NSE and the cluster-virial expansion,
we discuss the influence of the medium what is unavoidable to describe warm dense matter up to saturation density.
We consider the approximation of the EOS (\ref{eos}) where only the single-nucleon contributions are taken, i.e. the
sum over $A$ is reduced to $A=1$ which contains the neutron ($n$) and proton ($p$) quasiparticle contribution to the EOS.

In the quasi-particle approximation, the
imaginary part of $\Sigma$ is neglected in (\ref{spectral}). The spectral function
is $\delta$ like, and the densities are calculated from Fermi
distributions with the single-nucleon quasiparticle energies
$E_1(1) = \hbar^2p_1^2/2m + {\rm Re} \Sigma(1,E_1(1)) =  \hbar^2p_1^2/2m+ \Delta E^{\rm SE}(1)$
 so that (spin factor 2)
\begin{equation}
n^{\rm qu}_{\tau}(T,\mu_n,\mu_p)=\frac{2 }{ \Omega} \sum_{p} f_{1, \tau}[E_\tau(p;T,n_B,Y_p);T,\mu_n,\mu_p]\,.
\label{nqu}
\end{equation}

The quasiparticle approximation is well elaborated in nuclear physics, see \cite{Talmi,RS}. Starting from a microscopic
approach with suitable nucleon-nucleon interaction potentials, standard approximations for the 
single-nucleon self-energy shift $ \Delta E^{\rm SE}(1)$ are the Hartree-Fock-Bogoliubov or the Dirac-Brueckner-Hartree-Fock 
approximation, see Sec.~\ref{Sec:Introduction}. 
In the spirit of the density-functional approach, semi-empirical expressions such as the Skyrme forces
or relativistic mean-field approaches have been worked out. 
The relativistic quasiparticle energy
\begin{equation}
\label{DDRMF}
E_\tau(p;T,n_B,Y_p) = \sqrt{ \left[ m_\tau c^2-S(T,n_B,Y_p) \right]^2+\hbar^2 c^2 p^2} 
+ V_\tau(T,n_B,Y_p)- m_\tau c^2 
\end{equation}
gives in the non-relativistic limit $\Delta E^{\rm SE}_{\tau}(0)=-S(T,n_B,Y_p)+V_\tau(T,n_B,Y_p)$ 
and $m_\tau^*/m_\tau = 1-S(T,n_B,Y_p)/(m_\tau c^2)$. Explicit expressions for $ S(T,n_B,Y_p)$ and $V_\tau(T,n_B,Y_p) $ in
form of Pad\'e approximations
which are suitable for numerical applications, are given in Appendix \ref{app:1}. 
They are obtained from the DD-RMF parametrization of Typel \cite{Typel2005} and can be replaced by alternative parametrizations \cite{Providencia,Avancini,Hempel2014}.

Fitted to properties near the saturation density, the description of warm dense matter at densities near $n_{\rm sat}$ is adequate.
No cluster formation can be described in the single-nucleon quasiparticle (mean field) approach.
We have to go beyond this approximation and have to treat
 the imaginary part of $\Sigma$ in (\ref{spectral}) to include cluster formation and to reproduce the correct low-density limit.

\subsection{Shifts of light cluster binding energies in dense matter }
\label{sec:shiftsoflight}

Now we come back to the EOS (\ref{eos}) and add the contributions of clusters with $A=\{2,3,4\}$. We consider the bound state parts.
In the low-density limit we use the empirical binding energies $B_c=-E_c^0$, see below Tab. \ref{tablambda}, as also used in the NSE.
In the case of $c=\{d,t,h\}$ there is no excited bound state above the ground state.
In the case of $\alpha$, binding energy $B_\alpha$ = 28.3 MeV, there exists an excited $\alpha'$ state 
with excitation energy 20.2 MeV to be included 
into the partial density (\ref{zpart}) which also contains the contribution of scattering states.

Going to finite densities, the eigenvalues $E_{A,\nu}(P;T, n_B, Y_p,T_{\rm eff})$ 
(quasiparticle energies of the light clusters) depend, in the last consequence, on the temperature and chemical potentials
of the nuclear matter, as derived from the in-medium wave equation (\ref{waveA}).
The light clusters are considered as quasiparticles with dispersion relation depending on the 
single-nucleon occupation number $n(1)$ which is parametrized by a Fermi distribution (\ref{effparameter1})
with effective parameter values $\{T_{\rm eff},n_B,Y_p\}$.

We discuss the contribution to the dispersion relation of the cluster quasiparticles according to (\ref{qushift}).
The most significant medium effect is the Pauli blocking which is also strongly dependent on temperature.
The Pauli blocking shift of the binding energies $\Delta E_c^{\rm Pauli}(P;T_{\rm eff},n_B,Y_p)$, see Eq. (\ref{delpauli0P2}), 
has been evaluated within a variational approach.
The results are presented in  \cite{R2011}, 
and a parametrization has been given which allows to calculate the medium modification
of the bound state energies with simple expressions in good approximation, see Eq. (14) of Ref.  \cite{R2011}. 
We collect the results for the Pauli-blocking medium shifts of the bound state energies in App. \ref{App:qushift}.

The contribution of the single nucleon  energy shift to the cluster self-energy shift $\Delta E_{A,\nu}^{\rm SE}$ 
is easily calculated in the effective mass approximation, 
where the single-nucleon quasiparticle energy shift 
\begin{equation}
 \Delta E_\tau^{\rm SE}(p)=\Delta E_\tau^{\rm SE}(0)+\frac{\hbar^2p^2}{2m^*}-\frac{\hbar^2p^2}{2m}
\end{equation}
can be represented by the energy shift 
$\Delta E_\tau^{\rm SE}(p=0)$ and the effective mass $[m^*_\tau]^{-1}
=[m_\tau]^{-1}+\partial^2 \Delta E_\tau^{\rm SE}(p)/ \partial p^2|_{p=0}$.  
We use the empirical value 
\begin{equation}
	\frac{m^*}{m} = 1 -0.17 \frac{n_B}{n_{\rm sat}}.
\end{equation}

In the rigid shift approximation where $m^*=m$, the self-energy shift 
$\Delta E_\tau^{\rm SE}$ cancels in the binding energy because the continuum is shifted by the same value.																
It can be absorbed in the chemical potential of the EOS (\ref{quasigas}). 

In general, in the kinetic part of the the wave equation (\ref{waveAfree}) 
which consists of the single-nucleon quasiparticle energies $E_{\tau_i}(p_i;T,n_B,Y_p)$, we can introduce the c.o.m. momentum $\bf P$ 
and the intrinsic motion described by Jacobi coordinates.
In the effective mass approximation, the separation of the c.o.m motion is simple because the single-particle dispersion relations are quadratic.
 The  self-energy shift $\Delta E_{A,\nu}^{\rm SE}(P;T,n_B,Y_p)
=\Delta E_{A,\nu}^{\rm SE, c.o.m.}(P;T,n_B,Y_p)+\Delta E_{A,\nu}^{\rm SE, intr.}(P;T,n_B,Y_p)$
consists of the c.o.m. part $
 \Delta E_{A,\nu}^{\rm SE, c.o.m.}(P;T,n_B,Y_p)= E_{A,\nu}^{\rm cont}(P;T,n_B,Y_p)-\hbar^2 P^2/(2Am)$
which coincides with the edge of the continuum (\ref{Econt}) for the intrinsic motion,
 and the intrinsic part
\begin{equation}
\label{delkin}
 \Delta E_{A,\nu}^{\rm SE, intr.}(P;T,n_B,Y_p)=E_{A,\nu}^{\rm kin, intr.}\left(\frac{m}{m^*}-1\right).
\end{equation}
The intrinsic part of the cluster self-energy shift $\Delta E_{A,\nu}^{\rm SE, intr.}(P)$ 
is easily calculated for given wave functions \cite{SR,R}, see also \cite{Typel}, within perturbation theory.
Values for $E_{A,\nu}^{\rm kin, intr.}$ for the light elements are given below in Tab. \ref{Tabcd}.
It results as the averages of $\hbar^2/m q_1^2$ for $A=2$, $\hbar^2/m (q_1^2+3/4 q_2^2)$ for $A=3$, 
and $\hbar^2/m (q_1^2+3/4 q_2^3 +2/3 q_3^2)$ for $A=4$,
where ${\bf q}_i$ denote the respective Jacobian momenta \cite{R}.

We introduce the intrinsic part of the bound state energies as
\begin{eqnarray}
\label{Eint}
 E_{A,\nu}^{\rm intr.}(P;T,n_{B},Y_p,T_{\rm eff})&=&E_{A,\nu}(P;T,n_{B},Y_p,T_{\rm eff})-E_{A,\nu}^{\rm cont}(P;T,n_{B},Y_p)
\nonumber \\ &=&E_{A,\nu}^{0}+\Delta E_{A,\nu}^{\rm SE, intr.}(P;T,n_B,Y_p)+\Delta E_c^{\rm Pauli}(P;T_{\rm eff},n_B,Y_p)\,.
\end{eqnarray}
With Eq. (\ref{Ebind}), the intrinsic parts of the bound state energies are the negative values of the binding energies, 
$E_{A,\nu}^{\rm intr.}(P) \equiv -B_{A,\nu}^{\rm bind}(P)$.

\subsection{Mott points}
A consequence of the medium modification is the disappearance of bound states with increasing density 
what is of significance for the physical properties. To calculate the composition one has to check for given parameter values $\{T,n_B,Y_p\}$
whether the binding energy of the cluster with quantum numbers $\{A,\nu,{\bf P}\}$ is positive. We denote the density $n_{A,\nu}^{\rm Mott}(T,Y_p)$
as Mott density where the binding energy of a cluster $\{A,\nu\}$ with c.o.m. momentum ${\bf P}=0$ vanishes, with (\ref{Econt}), (\ref{Eint})
\begin{equation}
\label{nMott}
 E_{A,\nu}^{\rm intr.}(0;T,n_{A,\nu}^{\rm Mott},Y_p,T_{\rm eff}) =0\,.
\end{equation}
(Note that $T_{\rm eff}$ is determined by $\{T,n_B,Y_p\}$, see Eq. (\ref{Teff}) below.) 
For baryon densities $n_B>n_{A,\nu}^{\rm Mott}(T,Y_p)$ we can introduce
the Mott momentum ${\bf P}_{A,\nu}^{\rm Mott}(T,n_B,Y_p)$, where the bound state disappears, 
\begin{equation}
\label{PMott}
E_{A,\nu}^{\rm intr.}({\bf P}_{A,\nu}^{\rm Mott};T,n_{B},Y_p,T_{\rm eff})=0\,.
\end{equation}
At $n_B>n_{A,\nu}^{\rm Mott}(T,Y_p)$,
the summation over the momentum to calculate the bound state contribution to the composition is restricted to the region 
$|{\bf P}| >P_{A, \nu}^{\rm Mott}(T,n_B,Y_p)$. 

Crossing the Mott point by increasing the baryon density, part of correlations survive as continuum correlations 
so that the properties change smoothly. 
Therefore, the inclusion of correlations in the continuum is of interest.

\section{Virial expansion and correlated medium}
\label{sec:Virial}

In the low-density limit, rigorous expressions for the EOS are obtained for the virial expansion. 
The second virial coefficient is related 
to experimental data such as the bound state energies and scattering phase shifts, according to the Beth-Uhlenbeck formula \cite{Huang}.
The application to nuclear matter \cite{RMS,HS} as well as the generalized Beth-Uhlenbeck formula \cite{SRS} 
and the cluster-virial expansion \cite{Arcones2008,clustervirial} allow for the account of continuum correlations for the EOS. 

The virial coefficients are determined also by continuum correlations and are neglected in the simple NSE.
However, in particular for the deuteron contribution where the binding energy is small, the account for the 
correct second virial coefficient is of relevance, see the comparison of quantum statistical with generalized RMF
calculations in \cite{Typel}. 
A detailed description of the virial expansion in the context of a RMF treatment has been given by Voskresenskaya and Typel \cite{VT}. 
We are interested in the extension of the virial expansion to higher densities up to $n_{\rm sat}$.
For the two-nucleon case rigorous results can be given, whereas for the treatment of higher order correlations 
only some estimations can be made.

\subsection{Two-nucleon contribution}

The virial expansion of the EOS (\ref{eos}) reads \cite{RMS,SRS,HS,VT,Huang}
\begin{eqnarray}
&&n^{\rm tot}_n(T,\mu_n,\mu_p)=\frac{2}{\Lambda^3} \left[b_n(T) e^{\mu_n/T}+2b_{nn}(T)e^{2 \mu_n/T}+2 b_{np}(T) e^{(\mu_n+\mu_p)/T}
+\dots \right],
\nonumber \\ 
&&n^{\rm tot}_p(T,\mu_n,\mu_p)=\frac{2}{\Lambda^3} \left[b_p(T) e^{\mu_p/T}+2b_{pp}(T)e^{2 \mu_p/T}+2 b_{pn}(T) e^{(\mu_n+\mu_p)/T}
+\dots \right],
\end{eqnarray}

Already the noninteracting, i.e. ideal Fermi gas of nucleons contains two effects in contrast to the standard low-density, 
classical limit:\\
i) The relativistic dispersion relation $E_\tau( p)=c \sqrt{(m_\tau c)^2+ (\hbar c p)^2}-m_\tau c^2$ results in 
a first virial coefficient $b_\tau \neq 1$
where the value  $b_\tau = 1$ follows from the dispersion relation  $E_\tau( p)=(\hbar  p)^2/2 m_\tau $.
 For a more detailed investigation see \cite{VT}.\\
ii) The degeneration of the fermionic nucleon gas leads to the contribution $-2^{-5/2}$ to $b_{\tau \tau}$ \cite{Huang}.

The remaining part of  the second virial coefficient is determined by the two-nucleon interaction. We can introduce different channels, 
in particular the isospin triplet ($T_I=1$, neutron matter) and isospin singlet ($T_I=0$, deuteron)  channels 
which are connected with the spin singlet and spin triplet state, respectively, 
if even angular momentum is considered, e.g. S-wave scattering.
The second virial coefficient in both channels can be derived from $b_{nn}$ and $b_{np}$. 
Empirical values are given as function of $T$ in \cite{HS} (isospin symmetry is assumed).
\subsection{Generalized Beth-Uhlenbeck formula}
\label{GBU}
The second virial coefficients  $b_{nn}$ and $b_{np}$ cannot directly used within a quasiparticle approach.
Because part of the interaction is already taken into account when introducing the quasi-particle energy, 
we have to subtract this contribution from the second virial coefficient to avoid double counting, 
see \cite{clustervirial,SRS,VT}.
We expand the density in the quasiparticle approximation picture (\ref{nqu}), 
(\ref{DDRMF})  with respect to the fugacities. 
We identify the residual isospin-triplet contribution $v^0_{T_I=1}(T)$ from the neutron matter case as
\begin{eqnarray}
\label{n2viriala}
&&n^{\rm tot}_{B,\rm neutron\, m.}(T,\mu_n,\mu_p)=n^{\rm qu}_n(T,\mu_n,\mu_p)+\frac{2^{5/2}}{\Lambda^3}  e^{2\mu_n/T}v^0_{T_I=1}(T)+\dots,
\end{eqnarray}
and the residual isospin-singlet contribution $v^0_{T_I=0}(T)$ from the symmetric matter case
according to 
\begin{eqnarray}
\label{n2virial}&&n^{\rm tot}_{B,\rm symmetr. m.}(T,\mu_n,\mu_p)=n^{\rm qu}_n(T,\mu_n,\mu_p)+n^{\rm qu}_p(T,\mu_n,\mu_p)
\nonumber \\ &&+\frac{2^{5/2}3}{\Lambda^3} e^{(\mu_n+\mu_p)/T}
\left[e^{-E^0_d/T}-1+ v^0_{T_I=0}(T)+ v^0_{T_I=1}(T)+\dots \right],
\end{eqnarray}
dots indicate higher orders in densities.
The residual second virial coefficients $v^0_c(T)$ are given by \cite{SRS}
\begin{equation}
\label{resiv}
  v^0_c(T)=\frac{1}{\pi T}\int dE e^{-E/T} \left[\delta_c(E)-\frac{1}{2} \sin (2 \delta_c(E))\right]\,.
\end{equation}
Compared with the ordinary Beth-Uhlenbeck formula (\ref{B-U}) there are two differences:\\
i) After integration by parts, the derivative of the scattering phase shift is replaced by the phase shift, and according to the Levinson theorem 
for each bound state the contribution $-1$ appears. \\
ii) The contribution $-\frac{1}{2} \sin [2 \delta_c(E)]$ appears to avoid double counting when introducing the quasiparticle picture.

The EOS (\ref{eos}) is not free of ambiguities with respect to the subdivision into bound state contributions and continuum contributions, compare
(\ref{n2virial}), (\ref{resiv}) with (\ref{zpart}), (\ref{B-U}).  
The continuum correlations in $b_{\tau,\tau'}(T)$ are reduced to $v_c^0(T)$
if the quasiparticle picture is introduced.
The remaining contribution to the second virial coefficient $b_{\tau,\tau'}(T)$ is absorbed in the quasiparticle shift. 
This has been discussed in detail in \cite{SRS,clustervirial,VT}.

To give an approximation for $v_c^0(T)$, 
we performed  calculations within the generalized Beth-Uhlenbeck approach \cite{SRS} for a simple separable potential, 
\begin{equation}
\label{seppot}
 V_c(12,1'2')=-\lambda_c e^{-\frac{({\bf p}_1-{\bf p}_2)^2}{4\gamma^2}}e^{-\frac{({\bf p}'_1-{\bf p}'_2)^2}{4\gamma^2}}
 \delta_{\sigma,\sigma'}\delta_{\tau,\tau'}
\end{equation}
 with $\lambda_d= 1287.37$ MeV for the deuteron (isospin 0) channel, $\gamma = 1.474$ fm$^{-1}$, see \cite{R2011},
 adapted to binding energy and point rms radius of the deuteron. 
 After evaluating the T-matrix, the  scattering phase shifts are obtained, 
 and $v^0_d(T)$
 has been evaluated. For details see \cite{SRS}.
The result is approximated by
\begin{equation}
\label{vT0}
v^0_d(T)= v^0_{T_I=0}(T)\approx 0.30857+0.65327\,\, e^{-0.102424 \,\,T/{\rm MeV}}\,.
\end{equation}

A similar calculation has been performed for the isospin triplet channel. The empirical value for the $n-n$ scattering length (-18.818 fm) 
is reproduced
with the interaction potential (\ref{seppot}), $\lambda_{T_I=1} = 814.2$ MeV, leaving $\gamma$ unchanged. Also the  
effective range (2.834 fm) is well approximated. The resulting value $v^0_{T_I=1}(T)\approx 0.16$ is nearly independent of $T$.

Thus, for the EOS (\ref{eos}) we have the residual contribution of continuum correlations in the isospin-triplet channel 
as well as in the isospin-singlet channel. 
The contribution of the isospin-triplet channel $v^0_{T_I=1}(T)$ to the total baryon density is small and will be omitted in the present work.
In particular, applying the EOS to calculate the yields of the expanding fireball of heavy-ion collisions (HIC) within the freeze-out concept, 
the partial density of correlations in the isospin-triplet channel contributes to the free nucleon yields
because no stable bound state is formed in this channel. 

We take the residual continuum correlations in the isospin-singlet channel $v^0_{d}(T)$ into account for the EOS. 
Applying to HIC we assume that these continuum correlations can be added to the yield of deuterons for the expanding fireball.

 Note that the quasiparticle shift (\ref{DDRMF}) was introduced by fitting 
empirical values near the saturation density. The extrapolation to the low-density region is ambiguous. 
The empirical values for the second virial coefficient $b_{\tau, \tau'}(T)$ and the residual part $v^0_c(T)$ can be used 
to introduce the quasiparticle shift at low densities in a consistent way, but this problem is not subject of the present work.

\subsection{Contribution of higher clusters}

Quasiparticle approaches, such as the RMF approximation, can be considered as effective density functionals 
that contain correlations beyond the 
mean-field (Hartree-Fock-Bogoliubov) approximation. The remaining contribution of correlations in the continuum is treated 
as residual continuum contribution to the partial density  $z_{A,c}^{\rm part.}$ (\ref{components}) of the few-body channel $c$. 
We discussed two-nucleon correlations in the previous section. The question arises about the contribution of higher order correlations,
in particular, the role of $\alpha$ clustering. 
Although four-particle correlations are treated only in the fourth virial coefficient, 
$\alpha$-like correlations may become of significance for the EOS at low temperatures because of the high binding energy $B_\alpha$.
This particular feature of the low-density limit of the EOS is treated in the cluster-virial expansion \cite{clustervirial}, see also \cite{HS}.

A significant feature of few-body correlations is the formation of bound states. 
In addition to $d$, we discuss the few-body channels related to the light elements $t,\, h$, and $\alpha$. 
Higher clusters are also well-known as described by the nuclear data tables.
Note that at $A > 4$, also nearly bound states such as $^8$Be or $^5$He are of interest and should also be extracted 
from the continuum correlations.
The remaining residual continuum terms are assumed to give only small contributions to the EOS so that they can be omitted.
This is similar to the two-nucleon case where in the isospin-triplet channel the continuum correlations are mainly accounted for by the 
quasiparticle shift. However, in the present work, we focus on $A \leq 4$.

As discussed in the previous Sec. \ref{GBU}, the
contribution of bound states to the EOS is not simply given by the term $e^{-E^0_c/T}$ with the bound state energy $E^0_c$, 
but should be considered as integral over scattering phase shifts, see also \cite{Yokohama14}. 
The subdivision in the bound state contribution and the scattering state contribution to the density is not unique. 
Using partial integration and the Levinson theorem, the contribution $e^{-E^0_c/T}-1$ is obtained 
which remains smooth when the bound state energy goes to zero. The remaining contribution of continuum correlations 
is determined by the nucleon-nucleon interaction. In the two-nucleon case we found the expression $v^0_{T_I=0}(T)$
(\ref{n2virial}), (\ref{vT0}). The partial density $z_{d}^{\rm part.}(P;T,n_B,Y_p)$ (\ref{zpart}) of the $d$ channel
(we switched from the variables $\{T, \mu_n,\mu_p\}$ to the set $\{T,n_B,Y_p\}$)
reads in the zero-density limit
\begin{equation}
\label{intrd}
z_{d}^{\rm part.}(P;T,n_B=0,Y_p)= e^{(\mu_n+\mu_p)/T} \, e^{- \hbar^2 P^2/4mT}g_d\left[e^{-E^0_d/T}-1+ v^0_{T_I=0}(T)\right]
\end{equation}
which extends the contribution to the density from the bound state (energy below the continuum edge)
to the continuum of scattering states of the energy spectrum, see \cite{Yokohama14}.

Similar to the deuteron case, we assume that the expression 
\begin{equation}
\label{intrc}
z_{c}^{\rm part.}(P;T,n_B=0,Y_p)= e^{(N\mu_n+Z \mu_p)/T} \, e^{- \hbar^2 P^2/2AmT}g_c\left[e^{-E^0_c/T}-1+ v^0_{T_I=0}(T)\right]
\end{equation}
can be taken to continue the contribution of bound states to the continuum of scattering states
for the other light elements $t, h$. In the $\alpha$ case, there is also an excited state of the ground state, 
excitation energy $E_{\alpha^{'}}=E_\alpha+20.2$ MeV.
Within our estimations, the two bound states lead to larger continuum contribution,
\begin{equation}
\label{intra}
z_{\alpha}^{\rm part.}(P;T,n_B=0,Y_p)=e^{(2\mu_n+2 \mu_p)/T} e^{- \hbar^2 P^2/8mT}
\left[e^{-E^0_\alpha/T}+e^{-E^0_{\alpha^{'}}/T}-2+ 2\, v^0_{T_I=0}(T)\right] \,.
\end{equation}

To motivate the ansatz (\ref{intrc}), (\ref{intra}), 
we consider the effective interaction between the constituent nucleons of the cluster that are comparable with the nucleon-nucleon interaction in $d$. 
In the high-temperature region, continuum correlations are of relevance in the deuteron channel because of the small binding energy of 2.225 MeV.
The other cluster $\{t,h,\alpha\}$ are more strongly bound so that the contribution of continuum correlations is of less relevance for the EOS. 
Therefore we assume that the use of $v^0_{T_I=0}(T)$ for all residual virial coefficient $v_c^0(T)$ can be taken as rough estimation.

\subsection{Extrapolation to saturation densities, deuteron case}
Using the quasiparticle concept, the virial expansion introduced at low densities can be extended to arbitrary densities 
below the saturation density. 
Using Eq. (\ref{Eint}), the partial density (\ref{zpart}) of the channel $c$ (including $A$) with c.o.m. momentum $\bf P$ reads
\begin{eqnarray}
\label{resvirc}
&&z_{c}^{\rm part.}(P;T,n_{B},Y_p,T_{\rm eff})= e^{[N\mu_n+Z \mu_p-NE_n(P/A;T,n_B,Y_p)-ZE_p(P/A;T,n_B,Y_p)]/T} \nonumber\\&& \times
g_c \left\{\left[e^{-E_{c}^{\rm intr.}(P;T,n_{B},Y_p,T_{\rm eff})/T}-1\right] 
\Theta \left[-E_{c}^{\rm intr.}(P;T,n_{B},Y_p,T_{\rm eff})\right]+ v_{c}(P;T,n_B,Y_p)\right\}
\end{eqnarray}
if the set of variables $\{T,n_{B},Y_p\}$ is introduced. The residual continuum contributions 
$v_{c}(P;T,n_B,Y_p)$ are depending on the nucleon 
densities.

For $A=2$ the second virial coefficient has been investigated within a generalized Beth-Uhlenbeck approach starting from the quasiparticle approach \cite{SRS}. 
Not only the cluster binding energy is modified by self-energy shifts and Pauli blocking shifts, see Eqs. (\ref{waveA}), (\ref{qushift}). Also the in-medium 
scattering phase shifts are modified. The calculations for realistic nucleon-nucleon interaction \cite{SRS} show: \\
i) At the Mott point, the bound state disappears abruptly, but at the same time the scattering phase shifts jump by $\pi$ 
so that the total contribution (\ref{B-U}) to the virial coefficient changes smoothly.
In particular, the EOS which relates the total baryon density to the chemical potentials remains smooth. \\
ii) Near the saturation density, only the single-nucleon quasiparticle contribution to the density (\ref{eos}) remains. 
The correlated partial densities [$z_{A,c}^{\rm  part.}(P)$ with $A>1$] vanish
when the baryon density approaches the saturation density. 
However, part of correlations is also condensed in the quasiparticle approach. 
For instance, two-nucleon correlations are treated in the Brueckner approximation for the self-energy.

To estimate the density dependence of the residual virial coefficient $v_d(P;T,n_B,Y_p)$ (\ref{resiv}) we solved the in-medium wave equation (\ref{waveA})
for $A=2$ with the separable potential (\ref{seppot}). The Pauli blocking term was taken in the Tamm-Dancoff form $(1-f)(1-f)$ \cite{Po}, 
so that pairing has been neglected. 
The T-matrix has been solved taking into account the fermionic Pauli blocking terms. 
From the T-matrix, the in-medium scattering phase shifts are obtained. The contribution to the EOS (\ref{resiv}) according to the 
 generalized Beth-Uhlenbeck expression
  has been evaluated as function of $T$ and $n_B$ contained in the Pauli blocking terms (symmetric matter $Y_p=0.5$). 
  For details see also \cite{SRS}.
The result is approximated by
\begin{eqnarray}
\label{vshiftd}
&& v_d(P;T,n_B,Y_p)\approx \left[1.24+\left(\frac{1}{ v_{T_I=0}(T)} -1.24 \right) e^{\beta_d n_B/T}\right]^{-1}\,.
\end{eqnarray}
where $\beta_d =  1876.2 \, {\rm MeV \,\,fm}^3 $, and
$ v_{T_I=0}(T)$ given by Eq. (\ref{vT0}).
At high temperatures, most part of the $d$ component is due to the scattering states because the binding energy is small compared with the temperature,
and the residual virial coefficient describing continuum correlations is of relevance in the deuteron case. 
For the common treatment of the bound state and scattering state contribution see also Ref.~\cite{Yokohama14}. 
The dependence of $Y_p$ has been neglected.

We can interpret this result (\ref{resvirc}) as follows: The contribution $e^{-E_d(P;T,n_B,Y_p,T_{\rm eff})/T}-1$ of the  bound state ($d$) is decreasing with increasing density 
and disappears at the so-called Mott density.
The bound state merges with the continuum of scattering states and forms a resonance so that there remains a contribution to the baryon density.
When the density is further increasing, the resonance moves to higher energies and becomes broader. 
Consequently, the contribution to the continuum states is strongly ( $\approx$ exponentially) reduced. 

Note that to describe pairing at high densities and very low temperatures, the Tamm-Dancoff form $(1-f)(1-f)$ of the Pauli blocking used, for instance, 
in the Brueckner theory, must be replaced by the Feynman-Galitsky form $(1-f-f)$ according Eq.~(\ref{waveA}), see Ref. \cite{SRS}.

\subsection{Estimates for higher clusters $A=3,4$}

In contrast to the deuteron case $c \to d$, there is no simple way
 to estimate the continuum contribution of the other clusters with $A=3, 4$.  We take $Y_p=0.5$, neglect the dependence on $Y_p$,
 and estimate the residual virial contribution of the continuum $v_{c}(P;T,n_B,Y_p)$ at $P=0$.
 
All bound states behave quite similar, the binding energy is decreasing with increasing density.
The shift of the quasiparticle cluster bound state energies was considered elsewhere \cite{R2011}, see App.~\ref{App:qushift}. 
Similar to the deuteron case, we expect that also for the
heavier clusters a contribution of the continuum remains when the bound state is dissolved. To estimate this contribution, 
we consider a two-particle
system with an effective interaction of separable Gaussian type \cite{R2011} with fixed range parameter $\gamma$ 
but with effective coupling parameter 
$\lambda^{\rm eff}_c$ which reproduces the binding energy $B_c$ of cluster $c$. 
The corresponding parameter values are given in Tab. \ref{tablambda}.
Changing the strength $\lambda^{\rm eff}_c$, at the critical value $\lambda_c^{\rm crit}=885.996$ MeV, 
the bound state merges with the continuum. From the strict evaluation of the bound state (quasiparticle) energy \cite{R2011}, 
we know the so called Mott densities $n_{c}^{\rm Mott}(T,Y_p)$ (\ref{nMott})
where the bound states disappear, see Tab. \ref{tablambda}.
\begin{table}[htdp]
\caption{Effective coupling strengths}
\begin{center}
\begin{tabular}{|c|c|c|c|c|}
\hline
$c$ & $d$ & $t$ & $h$ & $\alpha$\\
\hline
$B_c$ [MeV] & 2.225 & 8.482 & 7.718 & 28.3\\
\hline
$\lambda^{\rm eff}_c$ [MeV] & 1287.37 & 1775.09 & 1724.3 & 2865\\
\hline
$n^{\rm Mott}_c(T=5)$ [fm$^{-3}$]& 0.00396527 & 0.00548355 & 0.00514539 & 0.00788889\\
$n^{\rm Mott}_c(T=10)$ [fm$^{-3}$] & 0.007987 & 0.0101218 & 0.0094643 & 0.0145317\\
$n^{\rm Mott}_c(T=15)$ [fm$^{-3}$] & 0.01197 & 0.015118 & 0.0141411 & 0.0211622\\
$n^{\rm Mott}_c(T=20)$ [fm$^{-3}$] & 0.0158613 & 0.0209388 & 0.0199201 & 0.0278919\\
\hline
$\beta_c/T(T=5)$ [fm$^{3}$]& 376.917 & 506.894 & 517.639 & 595.077\\
$\beta_c/T(T=10)$ [fm$^{3}$]& 187.62 & 274.613 & 281.421 & 323.049\\
$\beta_c/T(T=15)$ [fm$^{3}$]& 124.829 & 183.855 & 188.349 & 221.832\\
$\beta_c/T(T=20)$ [fm$^{3}$]& 94.2282 & 132.748 & 133.707 & 168.309\\
\hline
$\beta_c$ [MeV fm$^3$] &1876,2 & 2746.13 & 2814.2 & 3230.5\\
\hline
\end{tabular}
\end{center}
\label{tablambda}
\end{table}%

We calculated the relation between the coupling strength $\lambda_c$ of the separable potential (\ref{seppot}) 
and the residual virial $v_c(T)$ and found a simple relation $v_c(T) \propto \lambda_c^4$ in good approximation for all values $T$ under consideration.
On the other hand, we searched for an effective interaction strength $ \lambda_c(n_B) $ which mimics the shift of the binding energies, caused by 
the Pauli blocking as a density effect, by a reduced effective interaction strength $ \lambda_c(n_B) $ which, this way, becomes depending on $n_B$.
We use the ansatz $\lambda_c(n_B) = \lambda^{\rm eff}_c e^{-\beta_c\,n_B/4 }$, determine $ \beta_c$ from the known value $ \lambda^{\rm eff}_c  $ 
at zero density and the value $\lambda_c(n^{\rm Mott}_c)= \lambda_c^{\rm crit}$ at the Mott density $n^{\rm Mott}_c$ where the bound state disappear.

We get with (\ref{vT0})
\begin{eqnarray}
\label{vshiftc}
&& v_c(P;T,n_B,Y_p)\approx \left[1.24+\left(\frac{1}{ v_{T_I=0}(T)} -1.24 \right) e^{\beta_c n_B/T}\right]^{-1}\,.
\end{eqnarray}
as a fit formula for the residual virial coefficients.
Parameter values are found in Tab. \ref{tablambda}.

\subsection{Correlated medium and effective temperature}
\label{sec:effT}

As discussed in Sec. \ref{sec:cmf} it is obvious that not only the free nucleons are responsible for the in-medium
modifications of the quasiparticle properties, see Eqs. (\ref{mHF}), (\ref{mPb}). The nucleons found in clusters contribute to the mean field 
leading to the self-energy, 
but occupy also phase space and contribute to the Pauli blocking. 
The cluster mean-field approximation considers  few-body T matrices in the self-energy 
and in the kernel of the Bethe-Salpeter equation. It leads to similar expressions, 
see (\ref{effocc}),
but the free-nucleon Fermi distribution
$f_{1,\tau_1}(1)$ replaced by the effective occupation number (\ref{nnorm})
which contains also the distribution function $f_B(E_{B , \bar \nu}(  \bar P))$ for the abundance of the different 
bound states 
and the respective normalized wave functions $\psi_{B \bar \nu \bar P}(1 \dots B)$. 

The self-consistent determination of $n(1)$ for given $T, \mu_n, \mu_p$ is very cumbersome,
and we have to consider appropriate approximations.
The approximation where the occupation number distribution is normalized to the  total number of nucleons (\ref{moment1}) with given spin 
and isospin, $\sum_{p_1} n(1)=N^{\rm tot}_{\sigma_1,\tau_1}$ at $T_{\rm eff}=T$, i.e. a Fermi distribution function 
with $\{T,\mu_n^{\rm eff}, \mu_p^{\rm eff}\}$ where 
$\mu_n^{\rm eff}, \mu_p^{\rm eff}$ are determined by normalization to the total nucleon density, 
has been used in calculating the in-medium effects in \cite{SR,Typel}.
Formally, producing the in-medium effects, all nucleons 
in the medium are taken into account, but 
they are considered as uncorrelated, forming a Fermi distribution of free nucleons with temperature $T$.

A better approximation for the occupation number in momentum space  $n(1)$ is obtained if also a new parameter 
$ T_{\rm eff}$ is introduced to replace $T$.
As before, the normalization (\ref{moment1}) relates the phase space occupation $f_{1,\tau}( T_{\rm eff},n_B,Y_p)$ 
to the total nucleon densities $n_n^{\rm tot}, n_p^{\rm tot}$. 
In addition, one can consider the second moment (\ref{moment2}) of the phase space occupation to determine $T_{\rm eff}$.

From nuclear matter calculation it is well known that because of correlations the occupation in phase space 
is more diffuse compared with the ideal Fermi gas, and tails in the single-nucleon distribution 
are related to the nucleon-nucleon interaction. For instance, the phase space occupation function for various 
temperatures and densities was considered in \cite{ARSKK}.

At low densities, the phase space occupation is determined by the composition of the nuclear matter 
and the wave function of the corresponding clusters. In particular, at decreasing temperatures 
the abundance of $\alpha$ particles is increasing, and the occupation in phase space is determined by the 
internal momentum distribution of $\alpha$ particles.

Based on the results for the phase space occupation $n(1)$ obtained in \cite{ARSKK} for different values of $T$ and $n_B$,
we determined the effective temperature $ T_{\rm eff}(T,n_B,Y_p)$. For this, $f_{1,\tau}( T_{\rm eff},n_B,Y_p)$
was fitted to the calculated values of $n(1)$ so that not only the normalization $n_n^{\rm tot}, n_p^{\rm tot}$
are fulfilled, but also the maximum of the first derivative (at the Fermi energy) is reproduced.
A simple relation
\begin{equation}
\label{Teff}
 T_{\rm eff} \approx 5.5\, {\rm MeV} + 0.5\,\, T + 60\,\, n_B\,\, {\rm MeV\,\,fm}^3
\end{equation}
was obtained.

This means, we can use the parametrization of the cluster quasiparticle shifts as calculated for a phase space occupation
in the Pauli blocking which is described by a Fermi function. We have to consider the total density of 
all nucleons for the normalization. Furthermore, we have to replace  $T$ by $T_{\rm eff}$.

This estimation is only a simple fit that can be improved. We mention that at low  temperatures, 
$\alpha$ matter has to be described. Further effects like pairing are also not included here.
For the $\alpha$ correlations at very low temperature, quartetting \cite{PRL} is expected.
$\alpha$-like correlations in finite nuclei are also described at very low temperatures with the 
condensate  wave function, for infinite matter see \cite{Japaner}.
 For instance, as extreme case, we have to describe $\alpha$ matter 
 where the occupation of the phase space is given by the momentum distribution within the $\alpha$ particle, in particular at zero temperature.

\section{Results}
\label{Sec:Results}

We solve the equations (\ref{eos}) considering the different contribution with $A \leq 4$, i.e. besides the free nucleons $n,\,p$
also the channels related to $d,\,t,\,h,\,\alpha$. The intrinsic quantum number $\nu_c$ refers to the bound states as far as they exist,
and to the scattering states. The summation over $P$ is replaced by an integral, $ \frac{1}{\Omega} \sum_P \to \int d^3P/(2 \pi)^3$.

The main ingredient are the medium modified energies $E_{A,\nu}(P;T,n_B,Y_p;T_{\rm eff})$. 
They are determined as function of the total nucleon densities, 
\begin{equation}
\label{densities}
 n^{\rm tot}_n=(1-Y_p) n_B\,,\qquad n^{\rm tot}_p=Y_p n_B\,,
\end{equation}
and a further parameter $T_{\rm eff}(T,n^{\rm tot}_n,n^{\rm tot}_p)$, Eq. (\ref{Teff}), 
that takes the correlations in the medium into account when calculating the Pauli blocking effect. 
Consequently, equations (\ref{eos}) have the form
\begin{eqnarray}
\label{ntot}
n^{\rm tot}_n &=& \sum_{c=n,p,d,t,h,\alpha} N_c n_c(T,\mu_n,\mu_p;n^{\rm tot}_n,n^{\rm tot}_p,T_{\rm eff}), \nonumber \\
n^{\rm tot}_p &=& \sum_{c=n,p,d,t,h,\alpha} Z_c n_c(T,\mu_n,\mu_p;n^{\rm tot}_n,n^{\rm tot}_p,T_{\rm eff}).
\end{eqnarray}
For given $\{T,n^{\rm tot}_n,n^{\rm tot}_p\}$, a self-consistent solution of (\ref{ntot}) must be found which determines $\mu_n,\mu_p$.
Then, the EOS $\mu_n(T,n^{\rm tot}_n,n^{\rm tot}_p)$ and $\mu_p(T,n^{\rm tot}_n,n^{\rm tot}_p)$ are found.

More explicitly, Eqs. (\ref{ntot}) read $n^{\rm tot}_n=n_n+n_d+2 n_t+ n_h+2n_\alpha$ and
$n^{\rm tot}_p=n_p+n_d+n_t+2 n_h+2n_\alpha$.
The contribution of free neutrons and protons ($A=1$) to the total density is given by
\begin{equation}
 n_\tau(T,\mu_n,\mu_p;n^{\rm tot}_n,n^{\rm tot}_p,T_{\rm eff})=\frac{1}{\pi^2}\int_0^\infty dP \frac{P^2}{e^{[E_\tau(P;T,n_B, Y_p)-\mu_\tau]/T}+1}
\end{equation}
The single-nucleon quasiparticle energies $E_\tau(P;T,n_B, Y_p)$ are taken from a RMF approach (\ref{DDRMF}). 
A parametrization of a particular approach (DD-RMF, see \cite{Typel2005}) is given by Eqs. (\ref{scalar}), (\ref{vector}).

The contribution of the deuteron channel is (nondegenerated case)
\begin{eqnarray}
 n_d(T,\mu_n,\mu_p,n^{\rm tot}_n,n^{\rm tot}_p,T_{\rm eff})&=&\frac{3}{2 \pi^2}\int_0^\infty dP P^2 e^{[-E_n(P/2;T,n_B, Y_p)-E_p(P/2;T,n_B, Y_p)
 +\mu_n+\mu_p]/T}
 \\ &
 \times& \left\{\left[e^{-E^{\rm intr.}_d(P;T,n_B, Y_p,T_{\rm eff})/T}-1\right]
 \Theta[-E^{\rm intr.}_d(P;T,n_B, Y_p,T_{\rm eff})]+v_d(P;T,n_B,Y_p)\right\}\nonumber 
\end{eqnarray}
with the intrinsic in-medium bound state energy [negative binding energy $-B_d^{\rm bind}(P)$ (\ref{Ebind})], see Eq. (\ref{Eint}),
\begin{equation}
 E^{\rm intr.}_d(P;T,n_B, Y_p,T_{\rm eff})=E_d^0+\Delta E^{\rm SE,intr.}_d(P;T,n_B, Y_p)+\Delta E^{\rm Pauli}_d(P;T,n_B, Y_p,T_{\rm eff})\,.
\end{equation}
The contribution of the c.o.m. motion to the kinetic energy is given by 
$E_n(P/2;T,n_B, Y_p)+E_p(P/2;T,n_B, Y_p)$. 
Expressions for $\Delta E^{\rm SE,intr.}_d(P;T,n_B, Y_p)$ are given by (\ref{delkin}), and for $\Delta E^{\rm Pauli}_d(P;T_{\rm eff},n_B, Y_p)$ 
by (\ref{delpauli0P2}).
Beyond the Mott density $n_d^{\rm Mott}(T,Y_p)$, Eq. (\ref{nMott}), bound states arise only for c.o.m. momenta $P$ 
larger than the Mott momentum ${\bf P}^{\rm Mott}_d(T,n_B,Y_p)$, Eq. (\ref{PMott}). We must not solve these relations but use
 the $\Theta$ function which indicates the region where a bound state exists. 
The merge with the continuum is smooth because of the subtraction of 1. The intrinsic energy which is the difference between 
the bound state energy $E_d(P,T,n_B, Y_p,T_{\rm eff})$ and the edge of the continuum of scattering states 
$E_n(P/2;T,n_B, Y_p)+E_p(P/2;T,n_B, Y_p)$, see Eq. (\ref{Econt}), goes to zero
at the Mott point, and is compensated by the term -1. Above the Mott point, 
the residual virial contribution $v_d(P;T,n_B,Y_p)$ (\ref{vshiftd}) to the 
partial density in the deuteron channel remains. It is strongly decreasing with increasing density.

Similar expressions are also obtained for the other light elements contributing, as cluster states, to the density. In particular,
\begin{eqnarray}
 n_t(T,\mu_n,\mu_p,n^{\rm tot}_n,n^{\rm tot}_p,T_{\rm eff})&=&\frac{1}{\pi^2}
 \int_0^\infty dP P^2 e^{[-2 E_n(P/3;T,n_B, Y_p)-E_p(P/3;T,n_B, Y_p)+2 \mu_n+\mu_p]/T}
 \\ &&
 \times \left\{\left[e^{-E^{\rm intr.}_t(P;T,n_B, Y_p,T_{\rm eff})/T}-1\right]
 \Theta[-E^{\rm intr.}_t(P;T,n_B, Y_p,T_{\rm eff})]+v_t(P;T,n_B,Y_p)\right\}\,,\nonumber 
\end{eqnarray}
\begin{eqnarray}
 n_h(T,\mu_n,\mu_p,n^{\rm tot}_n,n^{\rm tot}_p,T_{\rm eff})&=&\frac{1}{\pi^2}
 \int_0^\infty dP P^2 e^{[-E_n(P/3;T,n_B, Y_p)-2 E_p(P/3;T,n_B, Y_p)+\mu_n+2 \mu_p]/T}
  \\ &&
 \times \left\{\left[e^{-E^{\rm intr.}_h(P;T,n_B, Y_p,T_{\rm eff})/T}-1\right]
 \Theta[-E^{\rm intr.}_h(P;T,n_B, Y_p,T_{\rm eff})]+v_h(P;T,n_B,Y_p)\right\}\,,\nonumber
\end{eqnarray}
\begin{eqnarray}
 n_\alpha(T,\mu_n,\mu_p,n^{\rm tot}_n,n^{\rm tot}_p,T_{\rm eff})&=&\frac{1}{2 \pi^2}\int_0^\infty dP P^2 
 e^{[-2 E_n(P/4;T,n_B, Y_p)-2E_p(P/4;T,n_B, Y_p)+2 
 \mu_n+2 \mu_p]/T}
 \\ &&
 \times \left\{\left[e^{-E^{\rm intr.}_\alpha(P;T,n_B, Y_p,T_{\rm eff})/T}-1\right]
 \Theta[-E^{\rm intr.}_\alpha(P;T,n_B, Y_p,T_{\rm eff})]\right. \nonumber \\ && \left.
 +\left[e^{-E^{\rm intr.}_{\alpha'}(P;T,n_B, Y_p,T_{\rm eff})/T}-1\right]
 \Theta[-E^{\rm intr.}_{\alpha'}(P;T,n_B, Y_p,T_{\rm eff})]+2 v_\alpha(P;T,n_B,Y_p)\right\}\,. \nonumber
\end{eqnarray}
The intrinsic in-medium bound state energy $E^{\rm intr.}_c$, see Eq. (\ref{Eint}), 
and the values for the shifts $\Delta E^{\rm SE,intr.}_c(P;T,n_B, Y_p)$ and 
$\Delta E^{\rm Pauli}_c(P;T,n_B, Y_p,T_{\rm eff})$ 
are given by Eqs. (\ref{delkin}), (\ref{delpauli0P2}). For the $\alpha$-like contribution, the excited state at $E^0_{\alpha'} = - 8.1$ MeV 
has been taken into account, with shifts estimated
by the values of the shifts for the ground state at $E^0_{\alpha} = - 28.3$ MeV. 
The residual virial contribution of continuum states $v_c(P;T,n_B,Y_p)$ is 
estimated by Eq. (\ref{vshiftc}).

\begin{figure}[htp] 
\begin{center}
 \includegraphics[width=15cm]{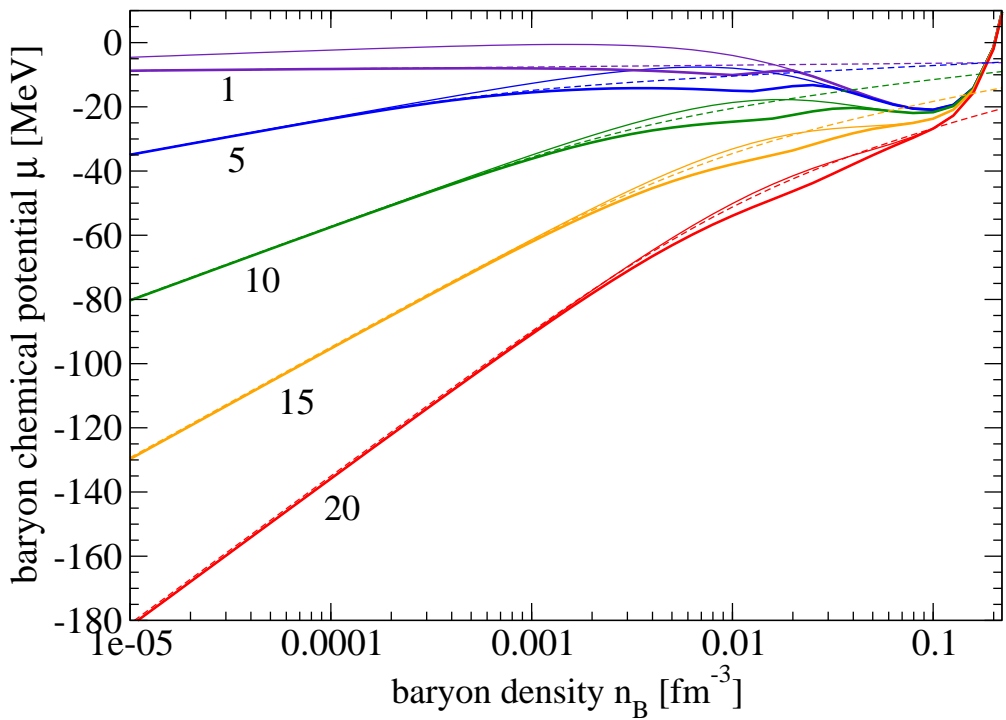}
\end{center}
  \caption{(Color online) Baryon chemical potential $\bar \mu$ as function of the baryon density $n_B$ for symmetric matter ($Y_p=0.5)$. Isotherms
  are shown for $T= 20, 15, 10, 5, 1$ MeV. The QS result (full line) is compared with the RMF solution (thin line) and the NSE solution (dashed).}
\label{fig_1}
\end{figure}

\subsection{EOS and critical point}
The self-consistent solution of Eqs. (\ref{ntot}) is shown in Fig. \ref{fig_1} ($n_B=n^{\rm tot}_n+n^{\rm tot}_p, Y_p=n^{\rm tot}_p/n_B$). 
Isotherms of the chemical potential 
$\bar \mu=g =(1-Y_p)\mu_n+Y_p \mu_p$ are shown for fixed asymmetry $Y_p=0.5$ (symmetric matter) as function of the baryon density $n_B$.
Temperatures are 20, 15, 10, and 5 MeV. The solution for $T=1$ MeV is shown for discussion, but in this case the formation of larger cluster 
is of importance.

For comparison, the pure mean-field (RMF)
solution (\ref{quasigas}) neglecting all contributions $A > 1$ is also shown. It dominates at low densities where (because of entropy) all 
bound states dissociate,
but becomes a good approximation at high densities where all bound states are blocked out and dissolved. 
The thermodynamics of the RMF approximation is 
modified in the region where clusters are formed. The chemical potential is lowered when correlations are taken into account.

The region where the mean-field approach is not sufficient depends strongly on $T$. 
For $T=20$ MeV deviations due to cluster formation appear below $n_B=0.07$ fm$^{-3}$, for $T=5$ MeV below $n_B=0.03$ fm$^{-3}$.
At the low-density region, the deviation from the mean-field solution (which coincides approximately with the free nucleon solution) 
is due to cluster formation as described by the mass-action law (NSE). It is well understood and also strongly depending on $T$.
At low temperatures (compare $T=1$ MeV) clustering occurs already at very low densities.

Also the NSE is shown what gives the correct behavior in the low-density limit. 
Cluster formation is described by the NSE, but deviations from $\bar \mu$ are 
shown as soon as the mean-field effects arise at about $n_B=10^{-3}$ fm$^{-3}$.

Thermodynamic stability requires $\partial \bar \mu / \partial n_B \geq 0$. 
As seen in Fig. \ref{fig_1}, below a critical temperature a phase transition appears, and a Maxwell construction can be applied.
For the pure  mean-field (RMF) solution (\ref{quasigas}) neglecting all contributions $A > 1$, 
the critical point is at $T_{\rm cr.}^{\rm RMF}=13.72$ MeV, $n_{B,{\rm cr.}}^{\rm RMF}=0.0486$ fm$^{-3}$.
Taking clustering with $A\leq 4$ into account, 
our QS approach gives $T_{\rm cr.}^{\rm QS}=12.42$ MeV, $n_{B,{\rm cr.}}^{\rm QS}=0.063$ fm$^{-3}$, see Tab. \ref{tab:Tc}.
The lowering of the critical temperature if clustering is taken into account is a general feature of many-particle systems.
The lowering of $T_{\rm cr.}$ due to clustering has been obtained for the QS approach in \cite{RMS} and, 
in contrast to the generalized RMF, also in \cite{Typel}. 
Calculations with the Skyrme interaction have been performed some time ago \cite{RMS2}, see also \cite{shenTeige,Vergleich}.

 \begin{table}[ht]
 \begin{tabular}{|c|c|c|c|c|c|}
 \hline
 & DD-RMF	& QS		& Skyrme \cite{RMS2}		&  Skyrme + $d$ \cite{RMS2}		& Skyrme + light \cite{RMS2}\\
 \hline
 critical temperature $T_{\rm cr.}$  [MeV]	& 13.72  & 12.42 & 22.7    & $21.1$   & 20.3\\
 critical density $n_{B,{\rm cr.}}$ [fm$^{-3}$]	& 0.0486    & 0.063   &-   & -    & -\\
 \hline
 \end{tabular}
 \caption{\label{tab:Tc}%
 Critical points from different approaches. DD-RMF and Skyrme: no clustering; QS: including light elements; + $d$: including deuterons;
 + light: including light elements.}
 \end{table}

A region of metastability is seen for low temperatures near $n_B \approx 0.02$ fm$^{-3}$. Note that the results at very low temperatures have
to be improved taking into account quantum condensates like pairing and quartetting. This is also possible within the QS approach,
introducing, e.g., the pair amplitude and performing a Bogoliubov transformation to new quasiparticles. 
For some results see \cite{PRL,PRL2,Japaner}. At low temperatures, also cluster with $A>4$ have to be taken into account.

\begin{figure}[htp] 
\begin{center}
a)
 \includegraphics[width=8cm]{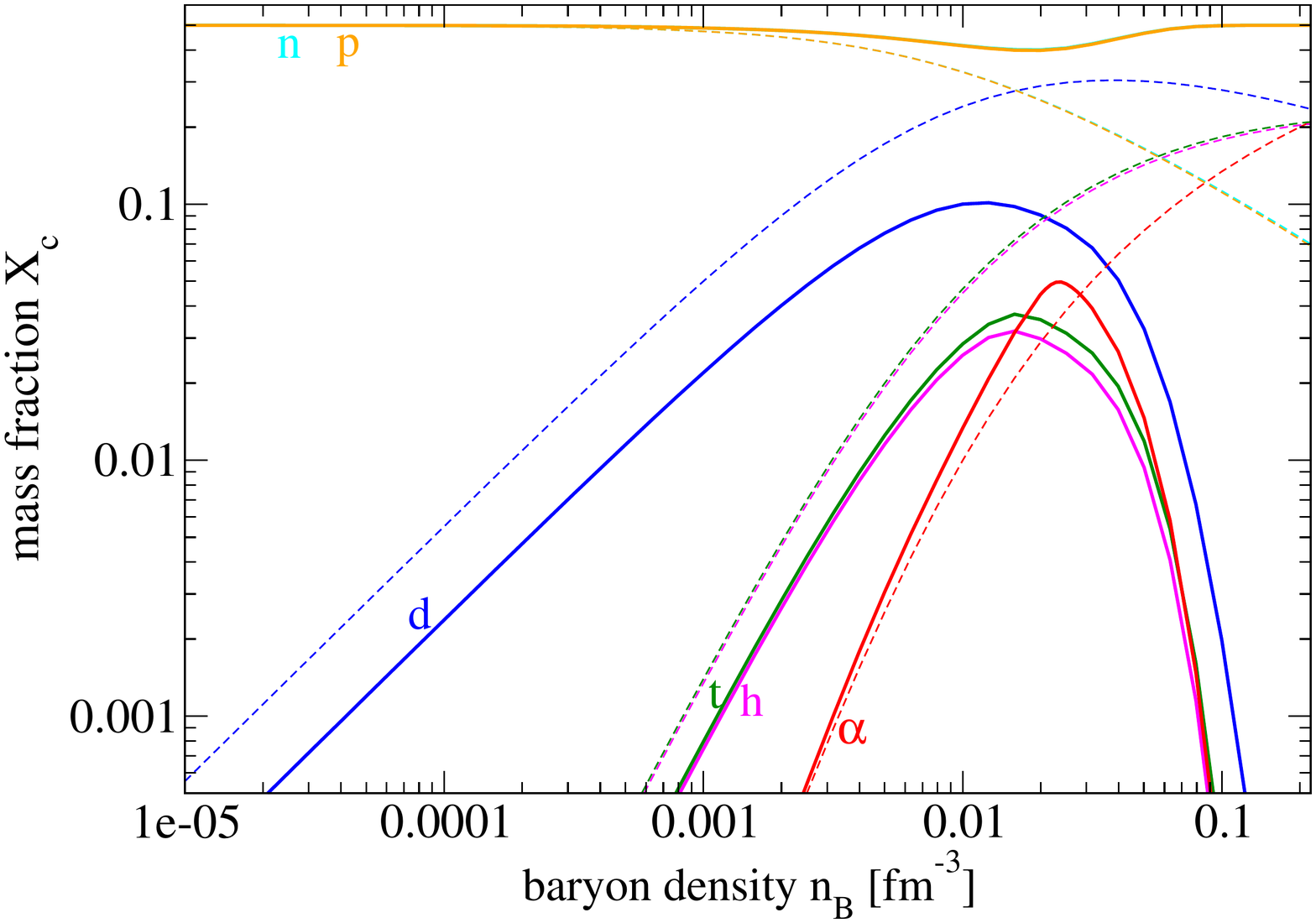}
 b)
 \includegraphics[width=8cm]{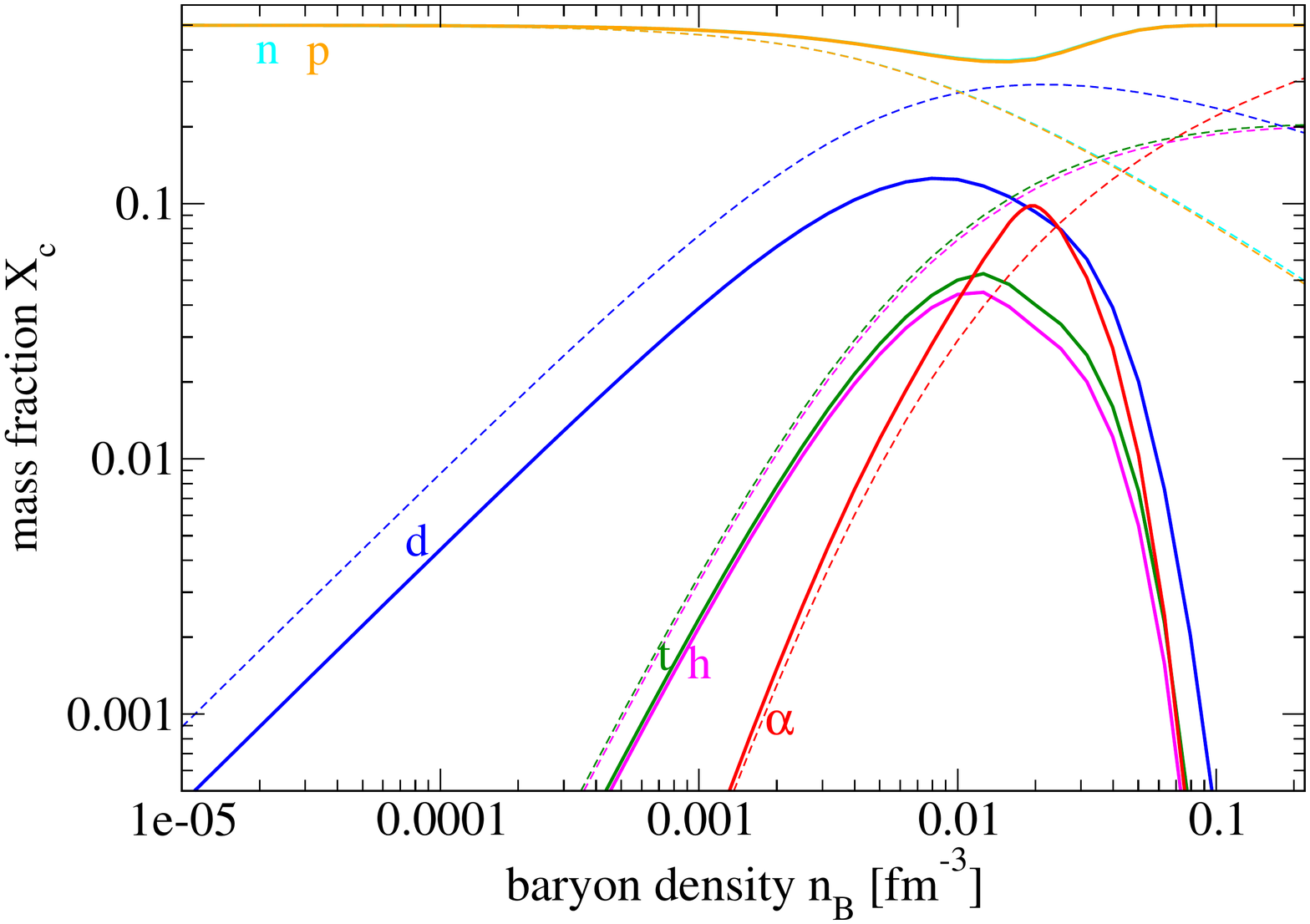}
\end{center}
  \caption{(Color online) Composition of symmetric matter ($Y_p=0.5$) at $T$ = 20 MeV [a)] and $T$ = 15 MeV [b)]. The mass fraction $X_c$
  is shown as function of the density $n_B$. The QS solution (full line) is compared with the NSE solution (dashed).}
\label{fig_2}
\end{figure}

\subsection{Composition above the critical point}

Of interest is the composition, i.e. the mass fractions $X_c= A_c n_c/n_B$. 
They are shown for symmetric matter ($Y_p=0.5$) at different temperatures 
as function of the baryon density $n_B$ in Fig. \ref{fig_2}, $c=\{n,p,d,t,h,\alpha\}$. 
The low-density region follows the mass action law (NSE), whereas at 
higher densities medium effects become of relevance and due to Pauli blocking all clusters are dissolved.
Whereas within the NSE the mass fraction $X_n,\,X_p$ of the free nucleons is continuously decreasing with density,
because of Pauli blocking the single quasiparticle mass fraction increases with density for $n_B>0.02$ fm$^{-3}$.
Correspondingly, the mass fractions of the cluster are strongly decreasing at high densities.

At low densities, the discrepancies between the QS and NSE results are due to the partial densities (\ref{zpart}) which contains 
also the continuum correlations. Compared with the simple NSE, the QS mass fractions $X_d,X_t,X_h$ are reduced because of the 
contribution $-1+v_c(T)$, see Eqs. (\ref{n2virial}), (\ref{intrc}). Thus, the contribution of the cluster to the total density
is lesser than expected from the simple NSE approach. For $X_\alpha$, the QS mass fraction is larger than the NSE result because 
the excited state $\alpha'$ is also taken into account (\ref{intra}).

%
\begin{figure}[htp] 
\begin{center}
a) \includegraphics[width=8cm]{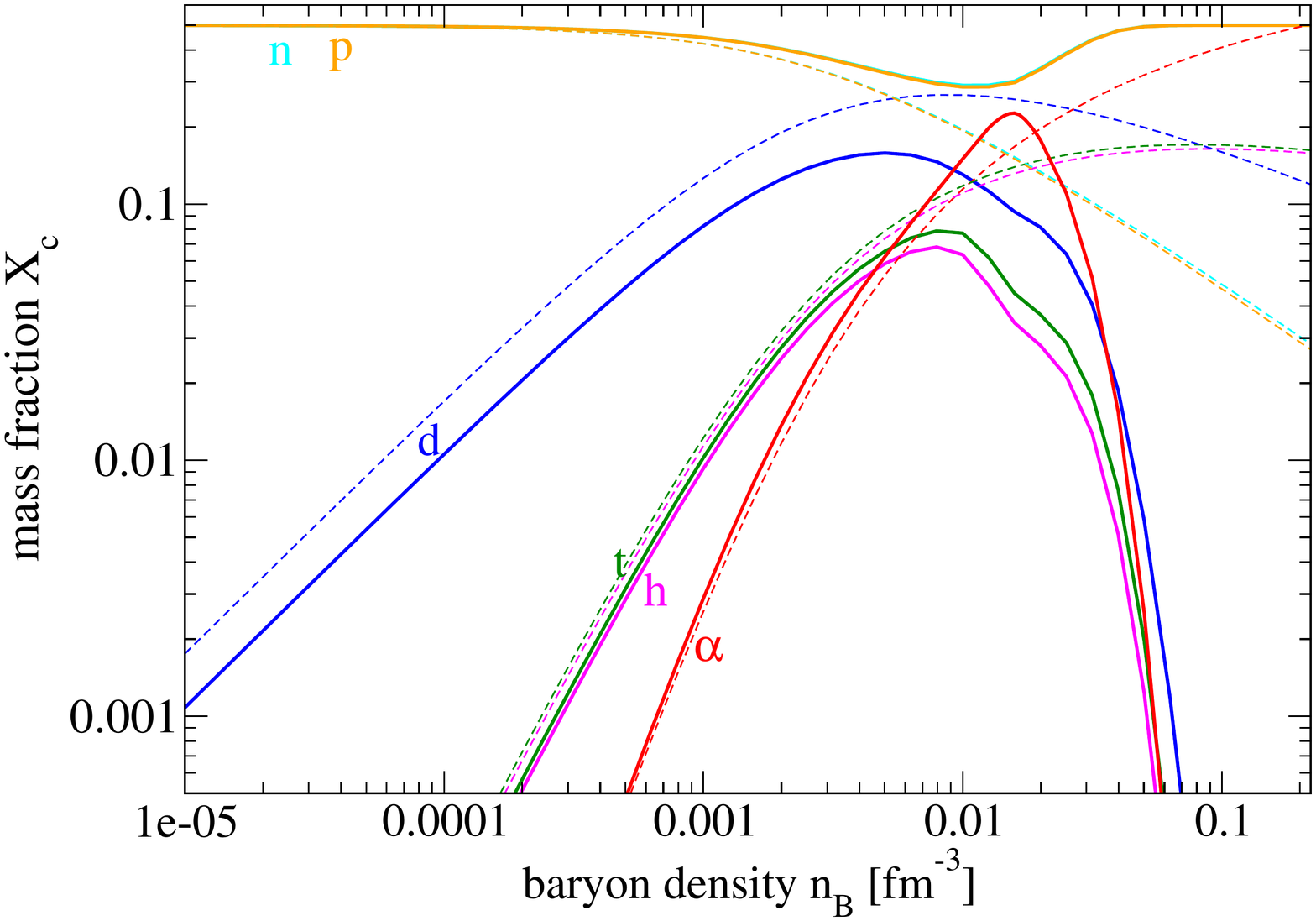}
b)  \includegraphics[width=8cm]{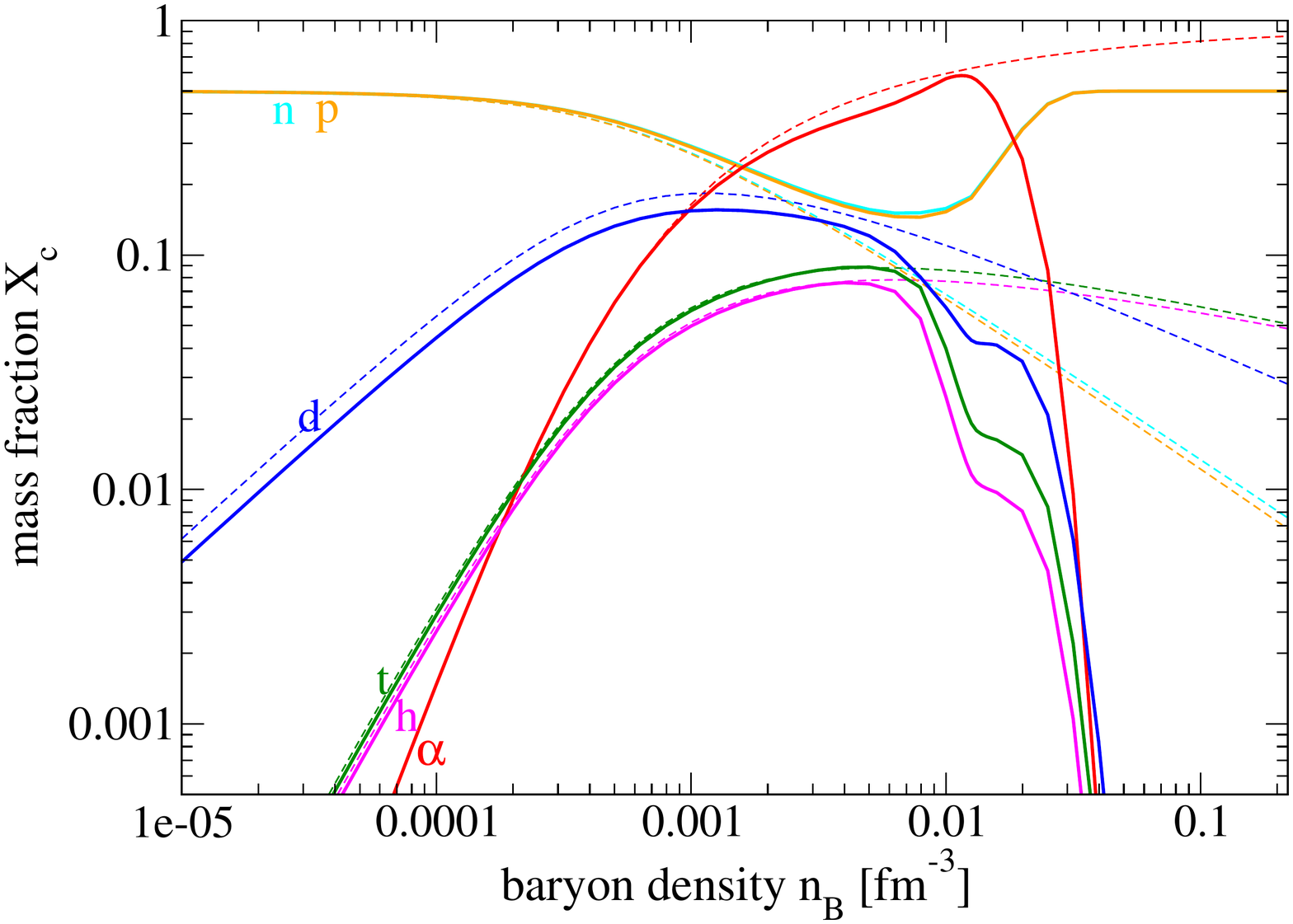}
\end{center}
  \caption{(Color online) Composition of symmetric matter ($Y_p=0.5$) at $T$ = 10 MeV [a)] and $T$ = 5 MeV [b)]. The mass fraction $X_c$
  is shown as function of the density $n_B$. The QS solution (full line) is compared with the NSE solution (dashed).}
\label{fig_4}
\end{figure}

\subsection{Composition below the critical point}

At lower temperatures, the role of correlations and cluster formation is increasing, see Fig.~\ref{fig_4}.
Whereas above $T_{\rm cr.}$ the two-particle correlations dominate, heavier cluster,
in particular $\alpha$-like correlations, give an increasing contribution to the composition
of warm dense matter. For instance, at $T=5$ MeV the mass fraction $X_\alpha$ is large in the density range $0.001 < n_B < 0.03$ [fm$^{-3}$].

In this intermediate density region, heavier clusters $A>4$ may be formed that are not included in the present work.
Moreover, the thermodynamic instability in that region leads to a first order phase transition of the liquid $\leftrightarrow$ gas type. 
Droplet formation and formation of pasta states may occur, see \cite{Furusawa,Avancini,Vergleich}. The inclusion of Coulomb interaction is indispensable, 
but we leave the theory of heavier cluster and phase transition out in this work.

\begin{figure}[htp] 
\begin{center}
 \includegraphics[width=8cm]{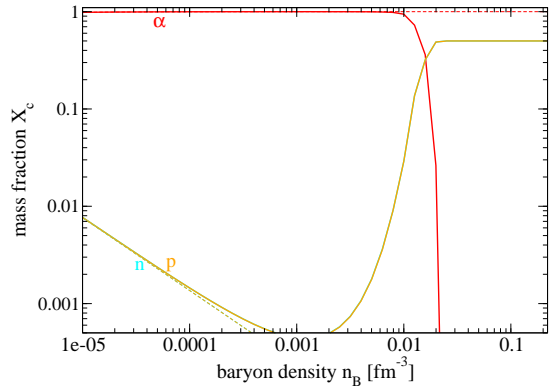}
\end{center}
  \caption{(Color online) Composition of symmetric matter ($Y_p=0.5$) at $T$ = 1 MeV. The mass fraction $X_c$
  is shown as function of the density $n_B$. The QS solution (full line) is compared with the NSE solution (dashed). 
  The mass fractions of $t,h, \alpha$ are very small and are not seen here.}
\label{fig_6}
\end{figure}

For illustration, also the case  $T=1$ MeV is considered in Fig.~\ref{fig_6}. Up to densities of $n_B \approx 0.02$ fm$^{-3}$,
$\alpha$-like correlations dominate and are dissolved due to the Pauli blocking terms. At higher densities, 
neutrons and protons as single-nucleon quasiparticles will describe the internal structure, well-described by RMF and 
other approaches. At very low temperatures, quantum condensates \cite{PRL,PRL2} appear which are not considered in the present work.

%


%
%

\section{Discussion}
\label{Sec:Discussion}

The present work intends to derive the EOS for warm dense matter in the entire subsaturation region, taking into account the 
known low-density virial expansions as well as the mean-field approaches near saturation density. Within a systematic 
quantum statistical approach, the single-nucleon quasiparticle description is improved by including few-body ($A\leq4$) 
correlations.
The QS approach gives a many-particle approach to describe self-energy effects, Pauli blocking, and correlations 
in the continuum. In the low-density region, the rigorous results of the virial expansion are reproduced, and the challenging
aspect is to use a generalized Beth-Uhlenbeck approach that is based on a quasiparticle approach, valid also near the saturation density. 
For the disappearance 
of bound states and residual virial coefficients in the density region $0.03 \leq n_B \leq 0.08$ [fm$^{-3}$] 
(where rigorous solution are not known) the effects continuum correlations are estimated.

The nuclear matter EOS (\ref{eos}) can be used to evaluate also other thermodynamic variables, including 
pressure, internal energy, entropy, symmetry energy, see App. \ref{App:TD}.
These thermodynamic variables are of interest in astrophysics where warm dense matter can occur. 
In particular, the physics of supernova core collapse enters the parameter region where cluster formation
with $A\leq 4$ in the subsaturation region is of relevance, and the evolution  (neutrino transport) is
determined by the presence of clusters. Whereas former approaches \cite{LS,Shen} considered only $\alpha$ particle 
formation, recently also other light elements are taken into account, within a quantum statistical model \cite{SR} 
or using the excluded volume concept \cite{Hempel}.

On the other hand, in heavy ion collision (HIC) a description is demanded which beyond NSE takes medium effects into account.
Recently, different versions of the EOS were compared with laboratory results \cite{Natowitz}. The QS approach which 
takes cluster formation $A\leq 4$ into account gives good agreement with the experimental data. Continuum correlations 
described by residual virial coefficients $v_c(P;T,n_B,Y_p)$ (\ref{vshiftc}) and the effective phase space occupation number 
$\tilde f_{1,\tau}(1;T_{\rm eff},n_B,Y_p)$ (\ref{effparameter1}) for the Pauli blocking energy shift have been included.
Furthermore, the symmetry energy \cite{Natowitz} is influenced by cluster formation in nuclear matter at low densities. 

Few-body correlations should also be considered 
in nuclear structure calculations. In particular, $\alpha$-like correlations appear in low-density isomers (e.g. the Hoyle state \cite{PRL2})
and in the low-density region at the surface of heavy nuclei that are $\alpha$ emitters (for instance $^{212}$Po \cite{Po}).
Here also a local density approach can be used to implement the results of the QS approach to the EOS.

We discuss some items to be improved in further works.

i) Input quantities are the RMF parametrization of the single-nucleon quasiparticle energy $E_{\tau}(P;T,n_B,Y_p)$ 
(\ref{selfe}), see App.~\ref{app:1},
the residual virial coefficients $v_c(P;T,n_B,Y_p)$, Eq. (\ref{vshiftc}), and the single-particle occupation number (\ref{effparameter})
parametrized by the total density of nucleons and the effective temperature (\ref{Teff}). Here, improvements are possible in future work.
This concerns also the parametrization of the cluster quasiparticle energies $E_{A\nu}(P;T,n_B,Y_p,T_{\rm eff})$ (\ref{qushift}), 
see App.~\ref{App:qushift} which has been given in previous works \cite{R2011}.

ii) A main disadvantage is the neglect of larger clusters ($A> 4$) what restricts the region of applicability of the present approach. 
A systematic QS approach to describe these correlations in warm dense matter is rather cumbersome, see \cite{Debrecen}. 
As a simple approach, the excluded volume may be introduced if we neglect the c.o.m. motion of the heavy nuclei.
Approaches \cite{Hempel} can be improved considering the intrinsic partition function. 
Then, the  region of phase instability can be treated, and Coulomb corrections are important. 
So-called nuclear pasta phases are discussed to derive an EOS also in the region of thermodynamic instability.

iii) It is a main problem to describe the transition from the gas phase state with well-defined clusters to the Fermi-liquid state. 
 The are some results at zero temperature where quantum condensates are formed. Pairing is not included in our approach because we use 
 the Tamm-Dancoff expression $(1-f)(1-f)$ instead of the Feynman-Galitsky expression $(1-f-f)$ for the Pauli blocking in the 
 Bethe-Salpeter equation (\ref{waveA}). Also quartetting \cite{PRL} is not described, but there are some
 results which consider the energy as function of density \cite{Japaner}. A related problem arises in nuclear structure \cite{Po,Yokohama14}.
 We expect that a more general and sophisticated approach to treat few-body correlations in warm dense matter will be worked out in future.

\begin{acknowledgments}
The author thanks D. Blaschke, T. Fischer, M. Hempel, K. Sumiyoshi, and S. Typel for many interesting discussions.

\end{acknowledgments}

\appendix

\section{Thermodynamic Potential for warm dense matter}
\label{App:TD}

In this work, we give solutions for the EOS (\ref{eos0}) for warm dense matter.
Note that there exist different EOS that refer to further thermodynamic variables like the pressure, the internal energy, or the entropy.
To get all thermodynamic quantities consistently, one can derive a thermodynamic potential. 
For instance, for fixed $Y_p$ the free energy  $F(T,\Omega,N_n,N_p)=\Omega f(T,n_B,Y_p)$ is found by integration, 
\begin{equation}
f(T,n_B,Y_p) =f(T,n_0,Y_p)+ \int_{n_0}^{n_B} \bar \mu(T, n',Y_p) d n'\,,
\end{equation}
where 
\begin{equation}
G/N_B=\bar \mu(T,n_B,Y_p)= (1-Y_p)\mu_n[T,(1-Y_P)n_B,Y_pn_B]+Y_p \mu_p [T,(1-Y_P)n_B,Y_pn_B]
\end{equation}
is the free enthalpy per baryon. 
$\mu_\tau(T,n^{\rm tot}_n,n^{\rm tot}_p)$ are the solutions of the EOS (\ref{eos}), $\tau=\{n,p\}$, 
and $\lim_{n_0\to 0}f(T,n_0,Y_p)$ follows from the solution of the free energy density for the ideal classical gas.

\section{The Cluster-mean field (CMF) approximation}
\label{App:CMF}

The chemical picture gives the motivation to extend the 
mean-field approximation for the case of cluster formation.
Bound states are considered as new species, to be treated on the same
level as free particles. A conserving mean-field approach can be formulated by specifying 
the Feynman diagrams that are taken into account when treating
the $A$-particle cluster propagation \cite{cmf}. The corresponding  
$A$-particle cluster self-energy is treated to first order in 
the interaction with the single particles ($n,p$) as well as with 
the $B$-particle cluster states ($d,t,h, \alpha$) in the medium, but with
full anti-symmetrization of the normalized wave functions of both clusters $A$ and $B$. We use the
notation $\{A,\nu,P\}$ for the particle number, internal quantum number (including proton number $Z$)
and center of mass momentum for the cluster under consideration and
$\{B,\bar \nu,\bar P\}$ for a cluster of the surrounding medium.

The Green function approach describes the propagation of a single nucleon by a Dyson equation governed by the self-energy,
and the few-particle states are obtained from a Bethe-Salpeter equation containing the effective interaction kernel.
For the $A$-particle
problem, the effective wave equation (\ref{eos}) can be rewritten as
\begin{eqnarray}
&&[E(1)+ \dots E(A) - E_{A ,\nu}( P)] \psi_{A \nu P}(1 \dots A)\nonumber\\
&& + \sum_{1'\dots A'}
\sum_{i<j}^A V_{ij}^A(1\dots A, 1'\dots A')  \psi_{A \nu P}(1' \dots A')
\nonumber\\ && + \sum_{1'\dots A'}
V_{\rm matter}^{A,{\rm mf}}(1\dots A, 1'\dots A')  \psi_{A \nu P}(1'
\dots A') = 0 \,,
\end{eqnarray}
with the kinetic energy $E(1)=\hbar^2 p_1^2/2 m_1$ and the interaction $V_{ij}^A(1\dots A, 1'\dots A') = V(12,1'2') \delta_{33'} \dots
\delta_{AA'}$.  The effective potential $V_{\rm matter}^{A,{\rm mf}}
(1\dots A, 1'\dots A')$ describes the influence of the nuclear
medium on the cluster bound states and has the form
\begin{equation}
V_{\rm matter}^{A,{\rm mf}}(1\dots A, 1'\dots A') = \sum_i^A \Delta  E^{\rm SE}
(i) \delta_{11'} \dots \delta_{AA'} + {\sum_{i,j}}' \Delta
V_{ij}^A(1\dots A, 1'\dots A') \,,
\end{equation}
with
\begin{eqnarray}
\label{cSE}
&&\Delta E^{\rm SE}(1) = \sum_2 V(12,12)_{\rm ex} n(2) -  \sum^\infty_{B=2}
\sum_{\bar \nu \bar P} \sum_{2 \dots B} \sum_{1' \dots B'} f_B[E_{B,  \bar \nu }( \bar P)]
\times \nonumber\\ && \qquad \qquad \qquad \times 
\sum_{i<j}^m V_{ij}^B(1\dots B, 1'\dots B') \psi_{B  \bar \nu  \bar
  P}(1 \dots B) 
 \psi^*_{B  \bar \nu  \bar P}(1' \dots B')\,,
\end{eqnarray}
\begin{eqnarray}
\label{cPb}
&&\Delta V^A_{12}(1\dots A, 1'\dots A') = - \Biggl\{\frac{1}{2}[n(1) + n(1')] 
V(12,1'2') +\\
&&\qquad \qquad +\sum_{B=2}^\infty \sum_{\bar \nu \bar P} \sum_{\bar 2 \dots \bar
  B} \sum_{\bar 
  2'\dots \bar B'} f_B[E_{B, \bar \nu}( \bar P)]   
\sum_j^B V_{1j}^B (1 \bar2' \dots \bar B', 1' \bar2 \dots
\bar B) 
\times  \nonumber\\ && \qquad \qquad \qquad \times 
\psi^*_{B \bar \nu \bar P} (2 \bar2 \dots \bar B) \psi_{B \bar \nu \bar P} (2' \bar2'
\dots \bar B') \Biggr\} \delta_{33'} \dots \delta_{AA'}\,, \nonumber
\end{eqnarray}
\begin{equation}
\label{effocc}
n(1) = f_1(1) + \sum^\infty_{B=2} \sum_{\nu \bar P} \sum_{2 \dots B}
B \,f_B[E_{B,  \bar \nu}(  \bar P)] |\psi_{B \bar \nu \bar P}(1 \dots B)|^2\,,
\end{equation}
where the variable $Z$ in the cluster distribution function (\ref{vert}) has not been given explicitly.

We note that within the mean-field approximation, the effective potential 
$V_{\rm matter}^{A,{\rm mf}}$ remains energy independent,
i.e.\ instantaneous.  The quantity $n(1)$ describes the effective
occupation of state $|1\rangle$ due to free and bound states, while exchange 
is included by the additional terms in $\Delta V^A_{12}$
and $\Delta(1)$, thus accounting for antisymmetrization.

A fully self-consistent solution of the cluster in a clustered
medium is a rather involved problem which has not been solved until
now. In particular, the composition of the medium has to be
determined, with energy shifts of the different
components (clusters of $B$ nucleons) in the medium
solving the effective wave equation for the $B$-nucleon problem.
The main contribution to the mean field is connected with the effective occupation number $n(1) $.
The exchange terms are necessary to describe quantum condensation
like pairing and quartetting.

\section{Nucleon quasiparticles}
\label{app:1}

The in-medium single-nucleon dispersion relation $E^{\rm qu}_{1}(P)$ can be expanded for small momenta $P$ as 
\begin{equation}
E^{\rm qu}_{1}(P)=  \frac{\hbar^2}{2 m_1}P^2 + \Delta E^{\rm SE}_{1}(P)=\Delta E^{\rm SE}_{1}(0) 
+ \frac{\hbar^2}{2 m_1^*}P^2 + {\mathcal O}(P^4)
\end{equation}
where the quasiparticle energies are shifted by $\Delta E^{\rm SE}_{1}(0)$, 
and $m_1^*$ denotes the effective mass of neutrons ($\tau_1=n$) or protons ($\tau_1=p$). 
Both quantities are functions  of $T,n^{\rm tot}_p,n^{\rm tot}_n$ characterizing the surrounding matter.

Different expressions are used to parametrize the quasiparticle shift in a large region of density, 
including nuclear saturation density. A standard approach to the nuclear matter EOS by Lattimer and Swesty \cite{LS} 
takes the Skyrme parametrization.

Alternatively, relativistic mean-field approaches have been developed starting from a model Lagrangian 
which couples the nucleons to mesons. A special parametrization, TM1, was used by Shen et al. \cite{Shen} 
to obtain the nucleon quasiparticle shift and the effective mass. 
Recent work on relativistic mean field approaches \cite{Typel2005} are fitted to reproduce properties of nuclei, 
but also are in agreement with microscopic DBHF calculations. 
A parametrization of the DD-RMF that is convenient for numerical calculations was given in \cite{clustervirial}:

For direct use, a parametrization for the DD
model \cite{Typel2005} was presented in Ref.\
\cite{Typel,clustervirial}.
We give here an improved parametrization of the DD2
  model \cite{clustervirial} in form of a Pad\'e approximation.
The variables are temperature $T$, baryon number
density $n_B=n^{\rm tot}_n+n^{\rm tot}_p$, and 
the asymmetry parameter  $\delta=1-2Y_p$  with the total proton fraction $Y_p=n^{\rm tot}_p/n_B$.
The intended relative accuracy in the parameter value range $T < 20$ MeV, $n_B< 0.16$ fm$^{-3}$ is 0.001.

The scalar self-energy (identical for neutrons and
  protons) is approximated as 
\begin{equation}
\label{scalar}
	S(T,n_B, \delta) = \frac{s_1(T,\delta) \;n_B + s_2(T,\delta) \;n_B^2
+  s_3(T,\delta)\;n_B^3}{1 + s_4(T,\delta)\;n_B + s_5(T,\delta)\; n_B^2}
\end{equation}
with coefficients
\begin{eqnarray}
 && s_i(T,\delta) = s_{i,0}(\delta) + s_{i,1}(\delta)\; T +  s_{i,2}(\delta)\;T^2  ,  \nonumber \\ 
 && s_{i,j}(\delta) = s_{i,j,0}+  s_{i,j,2}\;\delta^2 +  s_{i,j,4}\;\delta^4 ;
\end{eqnarray}
baryon number densities $n_B$ in fm$^{-3}$ and temperatures $T$ as well as the self-energies $S,V$ in MeV. Parameter
values are given in 
Table \ref{tab:S}.

\begin{table}[ht]
\begin{tabular}{|c|c|c|c|c|c|c|}
\hline
{} & $s_{i,j,k}$& $i=1$	& $i=2$		& $i=3$		& $i=4$		& $i=5$ \\
\hline
{}& $k=0$  & 4462.35	& 204334	&  125513	&  49.0026	&   241.935	\\
$j=0$& $k=2$	&  1.63811 &  -11043.9	&  -64680.5	&  -1.76282	&  -19.8568	 \\
{} & $k=4$  & 	0.293287&  -46439.7	 &  -4940.76	&  -10.6072	&  -48.3232	\\
\hline
{}& $k=0$  & -7.22458	&  7293.23	&   1055.3	&   1.70156	&   6.6665	\\
$j=1$& $k=2$	& 0.92618 & -49220.9	 &  -19422.6	&   -11.1142	&   -52.6306	 \\
{} & $k=4$  &-0.679133 	&  35263	&  15842.8	&   7.92604	&   38.1023	\\
\hline
{}& $k=0$  &  0.00975576  &	-209.452  &   132.502	&   -0.0456724	&   -0.112997	\\
$j=2$& $k=2$	&  -0.0355021 &	 2114.07  &  572.292	&   0.473553	&   2.15092	 \\
{} & $k=4$  & 	0.026292&  -1507.55	 &   -555.762	&   -0.337016	&   -1.57597	\\
\hline
\end{tabular}
\caption{\label{tab:S}%
Coefficients $s_{i,j,k}$ for the Pad\'e approximation of the scalar
  self-energy $S(T,n_B,\delta)$.}
\end{table}

The vector self-energy $V_p(T,n_B, \delta)=V_n(T,n_B, -\delta)$ is
approximated as
\begin{equation}
\label{vector}
	V_p(T,n_B, \delta) = \frac{v_1(T,\delta) \;n_B + v_2(T,\delta) \;n_B^2
+  v_3(T,\delta)\;n_B^3}{1 + v_4(T,\delta)\;n_B + v_5(T,\delta)\; n_B^2}
\end{equation}
with coefficients
\begin{eqnarray}
 && v_i(T,\delta) = v_{i,0}(\delta) + v_{i,1}(\delta)\; T +  v_{i,2}(\delta)\;T^2  ,  \nonumber \\ 
 && v_{i,j,k}(\delta) = v_{i,j,0}+  v_{i,j,1}\;\delta+  v_{i,j,2}\;\delta^2+  v_{i,j,3}\;\delta^3 +  v_{i,j,4}\;\delta^4 \,.
\end{eqnarray}
 Parameter
values are given in 
Table \ref{tab:V}.

\begin{table}[ht]
\begin{tabular}{|c|c|c|c|c|c|c|}
\hline
{} & $v_{i,j,k}$& $i=1$	& $i=2$		& $i=3$		& $i=4$		& $i=5$ \\
\hline
{}& $k=0$  & 3403.94		& -345.863 	&33553.8&2.7078		&18.7473	\\
{}& $k=1$  & -490.15		&  1521.62 	&4298.76&-0.162553	&4.0948364	\\
$j=0$& $k=2$ &  -0.0213143 	&  -2658.72	 &3692.23&-0.308454	&-0.0308012	\\
{}& $k=3$  & 0.00760759		&  -408.013	&-1083.14&-0.174442	&-0.751981	\\
{} & $k=4$  & 	0.0265109	&   -132.384	&-728.086&-0.0581052	&-0.585746	\\
\hline
{}& $k=0$  & -0.000978098	& 29.309	&-192.395&0.0161456	&-0.102959	\\
{}& $k=1$  & -0.000142646	&  -8.80748	&-52.0101&-0.00145171	&-0.044524	\\
$j=1$& $k=2$	&  0.00176929 	& -236.029	 &-141.702&-0.0689643	&-0.308021	\\
{}& $k=3$  & 0.00043752		& 13.7447	&-57.9237&-0.0000398794	&-0.0190921	\\
{} & $k=4$  & 	-0.00321724	&  111.538	&-11.4749&0.0317996	&0.0869529	\\
\hline
{}& $k=0$  & 0.0000651609	& 3.63322	&15.2158&0.00105179	&0.0118049	\\
{}& $k=1$  & 0.0000098168	& 0.0163495	&3.86652&0.000192765	&0.0021141	\\
$j=2$& $k=2$	&  -0.0000394036 & 6.88256 	 &-0.785201&0.00203728	&0.0070548	\\
{}& $k=3$  & 0.0000381407	&-0.369704	&1.59625&0.00000561467	&0.000565564	\\
{} & $k=4$  & 	0.000110931	& -3.28749	&2.0419	&-0.000932046	&-0.00182714	\\
\hline
\end{tabular}
\caption{\label{tab:V}%
Coefficients $v_{i,j,k}$ for the Pad\'e approximation of the vector
  self-energy $V_p(T,n_B,\delta)=V_n(T,n_B, -\delta)$.}
\end{table}

Note that these parametrizations are for the direct use, like the Skyrme or related models, based on some few 
input quantities, determined by empirical data. Various models, in particular RMF parametrizations, are presently
under discussion \cite{Providencia,Avancini,Hempel2014}.
It is the aim to give a good parametrization of the single-nucleon quasiparticle energies $E _\tau(P;T,n_B, Y_p)$,
as also well-known from density-functional approaches. The present RMF approach anticipates the dependence on $P$
according to (\ref{DDRMF}) but allows to adjust the dependence on $\{T,n_B,Y_p\}$.

\section{Shifts of bound state energies due to Pauli blocking}
\label{App:qushift}

The cluster quasiparticle energies have been calculated from Eq.~(\ref{waveAfree}) as function of $\{T, n^{\rm tot}_n, n^{\rm tot}_p\}$, 
 using a variational approach.
The single-nucleon occupation numbers are approximated by Fermi distributions $\tilde f_{1,\tau}(p;T_{\rm eff},n_B,Y_p)$ 
at the effective temperature $T_{\rm eff}$ and
normalized to the densities $n^{\rm tot}_n, n^{\rm tot}_p$.
 We consider the effect of Pauli blocking leading to the bound state energy shift 
 $\Delta E_{A,\nu}^{\rm Pauli}(P;T_{\rm eff},n_B,Y_p)$, Eq.~(\ref{qushift}). 
 For the calculations see Ref.~\cite{R2011}, where results for the Fermi function depending on $\{T,n_B,Y_p\}$ are derived. 
 We give the final expressions accordingly. 
 The variable $T$ has to be replaced by $T_{\rm eff}$ which now is relevant for the Pauli blocking expression.

The shifts of bound state energies due to Pauli blocking are approximated by
(we use $\nu=\{d,t,h,\alpha \}$ for the component $c$)
\begin{eqnarray}
\label{delpauli0P2}
&&\Delta E_\nu^{\rm Pauli}(P;T,n_B,Y_p)= 
c_{\nu}(P;T) \left\{1-\exp\left[- \frac{f_\nu(P;T,n_B)}{c_{\nu}(P;T)} y_\nu(Y_p)\,n_B- d_{\nu}(P;T,n_B)\,n_B^2 \right] \right\}\,.
\end{eqnarray}
The linear term $f_\nu(P;T,0)$ is given by first order perturbation theory 
with respect to the density, using the unperturbed wave 
functions of the free nuclei. Motivated by the 
exact solution for $A=2$ with the interaction potential (\ref{seppot}) (for details see \cite{R2011}) 
we use the following fit for arbitrary $n_B$ and $\nu$:
\begin{eqnarray}
\label{perturb}
&& f_\nu(P;T,n_B)= f_{\nu,1}  \exp\left[-\frac{P^2/\hbar^2}{4 ( f_{\nu,4}^2/f_{\nu,3}^2) (1+T/f_{\nu,2})+u_\nu n_B}\right] 
\frac{1}{T^{1/2}} \frac{2 f_{\nu,4}}{P/\hbar}   \nonumber\\
&&\times {\rm Im} \left\{ \exp\left[f_{\nu,3}^2 (1+f_{\nu,2}/T) \left(1-i\frac{P/\hbar}{2 f_{\nu,4}
(1+T/f_{\nu,2})}\right)^2\right] \right. \nonumber\\
&&\left. \times {\rm erfc}\left[ f_{\nu,3} (1+f_{\nu,2}/T)^{1/2} \left(1-i\frac{P/\hbar}{2 f_{\nu,4}
(1+T/f_{\nu,2})}\right)\right]\right\}
\end{eqnarray}
The parameter values $f_{\nu,i}$ and $u_\nu$ are given in Tab.~\ref{Tabcd}.

\begin{table}
\caption{ Parameter values for the Pauli blocking shift $\Delta E^{\rm Pauli}_\nu(P;T,n_B,Y_p)$, Eq.~(\ref{delpauli0P2}), 
for different nuclei ($\nu = \{d,t,h,\alpha$\}).}
\begin{center}
\hspace{0.5cm}
\begin{tabular}{|c|c|c|c|c|c|}
\hline
 parameter  &units&$d$ ($^2$H) & $t$ ($^3$H) & $h$ ($^3$He)&$\alpha$ ($^4$He) \\
\hline
 $ f_{\nu,1}$  &[MeV$^{3/2}$ fm$^3$] & 6792.6 & 20103.4 &19505.9 &36146.7\\
 $f_{\nu,2}$ &[MeV] &22.52 &11.987 & 11.748  &17.074  \\
 $f_{\nu,3}$ &-&0.2223 &0.85465  & 0.84473  &0.9865\\
  $f_{\nu,4}$&[fm$^{-1}]$&0.2317&0.9772&0.9566&1.9021\\
\hline
 $c_{\nu,0}$ & [MeV]& 2.752 &11.556&10.435 &150.71 \\
 $c_{\nu,1}$ &[ MeV$^3$]& 32.032  & 117.24 &176.78  &9772 \\
 $c_{\nu,2}$   &[ MeV] &  0 & 3.7362 &  3.5926 &2.0495 \\
$c_{\nu,3}$ & [MeV$^2$]&  9.733 & 4.8426 & 5.8137  &2.1624\\
\hline
$d_{\nu,1}$  & [MeV$^2$ fm$^6$] &  523757  & 108762 & 90996  &5391.2 \\
 $d_{\nu,2}$  & [MeV] & 0&  9.3312 &  10.72 &    3.5099 \\
 $d_{\nu,3}$ & [MeV$^2$ ] &15.273 &49.678 &47.919 & 44.126\\
\hline
$u_{\nu}$  & [fm ]&11.23  &25.27 & 25.27  &44.92 \\
\hline
$w_{\nu}$  & [MeV$^{-1}$ fm]& 0.145  &0.284 & 0.27  &0.433 \\
\hline
$E_\nu^{\rm kin, intr.}$  & [MeV]&10.338   &23.735 & 23.021  &51.575 \\
\hline
\end{tabular}
\label{Tabcd}
\end{center}
\end{table}

The dependence on the asymmetry $Y_p$ is also determined discussing the low-density limit.
For the deuteron, Pauli blocking is determined by the sum of the 
neutron and proton distribution functions $f_{1,n}(p_1)+f_{1,p}(p_1)$, see Eq.~(\ref{waveAfree}). 
It depends only on the total baryon density $n_B$ in the non-degenerate limit.
Therefore, we neglect the dependence on $Y_p$ for species $\nu \to d$. The same applies for $^4$He 
because neutron and proton orbitals are 
equally occupied for species $\nu \to \alpha$. In the clusters with $A=3$, however, 
neutrons and protons contribute differently to the internal structure, 
so that the shifts of $^3$H and $^3$He are sensitive to the asymmetry of nuclear matter. 
This leads to the expression $y_\nu(Y_p)$ in Eq.~(\ref{delpauli0P2})
with $y_d(Y_p)=y_\alpha(Y_p)=1$, for triton $y_t(Y_p)=\left( \frac{4}{3}-\frac{2}{3} Y_p\right)$, 
and for helion $y_h(Y_p)=\left(\frac{2}{3}+\frac{2}{3} Y_p\right)$.
For example, in comparison with helions ($^3$He), the tritons ($^3$H) show a stronger 
shift in neutron-rich matter because the neutrons in the cluster are stronger blocked than the protons. 

In the  fit formula (\ref{delpauli0P2}), the term
\begin{equation}
c_{\nu}(P;T)=c_{\nu}(0;T)=c_{\nu,0}+\frac{c_{\nu,1}}{(T-c_{\nu,2})^2+c_{\nu,3}}
\end{equation}
is not depending on $P$, but
\begin{equation}
 d_{\nu}(P;T,n_B)=d_{\nu}(0;T,n_B)\,\exp\left[-\frac{P^2/\hbar^2}{w_\nu T n_B}\right],\qquad 
 d_{\nu}(0;T,n_B)=\frac{d_{\nu,1}}{(T-d_{\nu,2})^2+d_{\nu,3}}, 
\end{equation}
is depending on $P$ with the parameter $w_{\nu}$.
The corresponding parameter values are given in Tab.~\ref{Tabcd}.

\end{document}